\input harvmac
\let\includefigures=\iftrue
\let\useblackboard=\iftrue
\newfam\black

\includefigures
\message{If you do not have epsf.tex (to include figures),}
\message{change the option at the top of the tex file.}
\input epsf
\def\figin{\epsfcheck\figin}\def\figins{\epsfcheck\figins}
\def\epsfcheck{\ifx\epsfbox\UnDeFiNeD
\message{(NO epsf.tex, FIGURES WILL BE IGNORED)}
\gdef\figin##1{\vskip2in}\gdef\figins##1{\hskip.5in}
\else\message{(FIGURES WILL BE INCLUDED)}%
\gdef\figin##1{##1}\gdef\figins##1{##1}\fi}
\def\DefWarn#1{}
\def\figinsert{\goodbreak\midinsert}
\def\ifig#1#2#3{\DefWarn#1\xdef#1{fig.~\the\figno}
\writedef{#1\leftbracket fig.\noexpand~\the\figno}%
\figinsert\figin{\centerline{#3}}\medskip\centerline{\vbox{
\baselineskip12pt\advance\hsize by -1truein
\noindent\footnotefont{\bf Fig.~\the\figno:} #2}}
\endinsert\global\advance\figno by1}
\else
\def\ifig#1#2#3{\xdef#1{fig.~\the\figno}
\writedef{#1\leftbracket fig.\noexpand~\the\figno}%
\global\advance\figno by1} \fi

\def\id{{1 \kern-.28em {\rm l}}}

\def\K3{{\bf K3}}
\def\journal#1&#2(#3){\unskip, \sl #1\ \bf #2 \rm(19#3) }
\def\andjournal#1&#2(#3){\sl #1~\bf #2 \rm (19#3) }

\def\bar{\overline}

\def\ie{{\it i.e.}}
\def\eg{{\it e.g.}}

\def\frac#1#2{{#1\over#2}}

\def\half{\frac12}

\def\inbar{\,\vrule height1.5ex width.4pt depth0pt}
\def\IC{\relax\hbox{$\inbar\kern-.3em{\rm C}$}}
\def\IR{\relax{\rm I\kern-.18em R}}
\def\IP{\relax{\rm I\kern-.18em P}}

%
%

%
\catcode`\@=11
\def\slash#1{\mathord{\mathpalette\c@ncel{#1}}}
\overfullrule=0pt

\def\EE{{\cal E}}

\def\MM{{\cal M}}
\def\NN{{\cal N}}

\def\underrel#1\over#2{\mathrel{\mathop{\kern\z@#1}\limits_{#2}}}

\catcode`\@=12


%

\def \sinh{{\rm sinh}}
\def \cosh{{\rm cosh}}

\def\exp{{\rm exp}}


\def\ie{{\it i.e.}}
\def\eg{{\it e.g.}}


\lref\KutasovUF{
  D.~Kutasov,
  ``Introduction to little string theory,''
ICTP Lect.\ Notes Ser.\  {\bf 7}, 165 (2002).
}

\lref\KutasovRR{
  D.~Kutasov,
  ``Accelerating branes and the string/black hole transition,''
[hep-th/0509170].
}

\lref\AharonyKS{
  O.~Aharony,
  ``A Brief review of 'little string theories',''
Class.\ Quant.\ Grav.\  {\bf 17}, 929 (2000).
[hep-th/9911147].
}

\lref\GiveonMI{
  A.~Giveon, D.~Kutasov, E.~Rabinovici and A.~Sever,
  ``Phases of quantum gravity in AdS(3) and linear dilaton backgrounds,''
Nucl.\ Phys.\ B {\bf 719}, 3 (2005).
[hep-th/0503121].
}

\lref\LarsenGE{
  F.~Larsen,
  ``A String model of black hole microstates,''
Phys.\ Rev.\ D {\bf 56}, 1005 (1997).
[hep-th/9702153].
}

\lref\SmirnovLQW{
  F.~A.~Smirnov and A.~B.~Zamolodchikov,
  ``On space of integrable quantum field theories,''
[arXiv:1608.05499 [hep-th]].
}

\lref\CavagliaODA{
  A.~Cavagliˆ, S.~Negro, I.~M.~Szcsnyi and R.~Tateo,
  ``$T \bar{T}$-deformed 2D Quantum Field Theories,''
[arXiv:1608.05534 [hep-th]].
}

\lref\GiveonNIE{
  A.~Giveon, N.~Itzhaki and D.~Kutasov,
  ``$T\bar T$ and LST,''
[arXiv:1701.05576 [hep-th]].
}

\lref\MaldacenaKY{
  J.~M.~Maldacena,
  ``Black holes in string theory,''
[hep-th/9607235].
}

\lref\CallebautOMT{
  N.~Callebaut, J.~Kruthoff and H.~Verlinde,
  ``$ T\overline{T} $ deformed CFT as a non-critical string,''
JHEP {\bf 2004}, 084 (2020).
[arXiv:1910.13578 [hep-th]].
}

\lref\HorneGN{
  J.~H.~Horne and G.~T.~Horowitz,
  ``Exact black string solutions in three-dimensions,''
Nucl.\ Phys.\ B {\bf 368}, 444 (1992).
[hep-th/9108001].
}

\lref\GiveonFU{
  A.~Giveon, M.~Porrati and E.~Rabinovici,
  ``Target space duality in string theory,''
Phys.\ Rept.\  {\bf 244}, 77 (1994).
[hep-th/9401139].
}

\lref\GiveonRW{
  A.~Giveon, A.~Konechny, E.~Rabinovici and A.~Sever,
  ``On thermodynamical properties of some coset CFT backgrounds,''
JHEP {\bf 0407}, 076 (2004).
[hep-th/0406131].
}

\lref\JohnsonQT{
  C.~V.~Johnson, A.~W.~Peet and J.~Polchinski,
  ``Gauge theory and the excision of repulson singularities,''
Phys.\ Rev.\ D {\bf 61}, 086001 (2000).
[hep-th/9911161].
}

\lref\MarolfND{
  D.~Marolf,
  ``The dangers of extremes,''
Gen.\ Rel.\ Grav.\  {\bf 42}, 2337 (2010).
[arXiv:1005.2999 [gr-qc]].
}

\lref\ChakrabortyMDF{
  S.~Chakraborty, A.~Giveon and D.~Kutasov,
  ``$T\bar{T}$, $J\bar{T}$, $T\bar{J}$ and String Theory,''
J.\ Phys.\ A {\bf 52}, no. 38, 384003 (2019).
[arXiv:1905.00051 [hep-th]].
}

\lref\GiveonZM{
  A.~Giveon, D.~Kutasov and O.~Pelc,
  ``Holography for noncritical superstrings,''
JHEP {\bf 9910}, 035 (1999).
[hep-th/9907178].
}

\lref\AharonyUB{
  O.~Aharony, M.~Berkooz, D.~Kutasov and N.~Seiberg,
  ``Linear dilatons, NS five-branes and holography,''
JHEP {\bf 9810}, 004 (1998).
[hep-th/9808149].
}

\lref\KutasovXU{
  D.~Kutasov and N.~Seiberg,
  ``More comments on string theory on AdS(3),''
JHEP {\bf 9904}, 008 (1999).
[hep-th/9903219].
}

\lref\AsratTZD{
  M.~Asrat, A.~Giveon, N.~Itzhaki and D.~Kutasov,
  ``Holography Beyond AdS,''
Nucl.\ Phys.\ B {\bf 932}, 241 (2018).
[arXiv:1711.02690 [hep-th]].
}

\lref\ChakrabortyKPR{
  S.~Chakraborty, A.~Giveon, N.~Itzhaki and D.~Kutasov,
  ``Entanglement beyond AdS,''
Nucl.\ Phys.\ B {\bf 935}, 290 (2018).
[arXiv:1805.06286 [hep-th]].
}

\lref\AharonyICS{
  O.~Aharony, S.~Datta, A.~Giveon, Y.~Jiang and D.~Kutasov,
  ``Modular covariance and uniqueness of $J\bar{T}$ deformed CFTs,''
JHEP {\bf 1901}, 085 (2019).
[arXiv:1808.08978 [hep-th]].
}

\lref\ChakrabortyVJA{
  S.~Chakraborty, A.~Giveon and D.~Kutasov,
  ``$ J\overline{T} $ deformed CFT$_{2}$ and string theory,''
JHEP {\bf 1810}, 057 (2018).
[arXiv:1806.09667 [hep-th]].
}

\lref\ChakrabortyMDF{
  S.~Chakraborty, A.~Giveon and D.~Kutasov,
  ``$T\bar{T}$, $J\bar{T}$, $T\bar{J}$ and String Theory,''
J.\ Phys.\ A {\bf 52}, no. 38, 384003 (2019).
[arXiv:1905.00051 [hep-th]].
}

\lref\HashimotoWCT{
  A.~Hashimoto and D.~Kutasov,
  ``$ T\overline{T},J\overline{T},T\overline{J} $ partition sums from string theory,''
JHEP {\bf 2002}, 080 (2020).
[arXiv:1907.07221 [hep-th]].
}

\lref\HashimotoHQO{
  A.~Hashimoto and D.~Kutasov,
  ``Strings, Symmetric Products, $T \bar{T}$ deformations and Hecke Operators,''
[arXiv:1909.11118 [hep-th]].
}

\lref\ApoloQPQ{
  L.~Apolo and W.~Song,
  ``Strings on warped AdS$_{3}$ via $ T\bar{J} $ deformations,''
JHEP {\bf 1810}, 165 (2018).
[arXiv:1806.10127 [hep-th]].
}

\lref\AharonyBAD{
  O.~Aharony, S.~Datta, A.~Giveon, Y.~Jiang and D.~Kutasov,
  ``Modular invariance and uniqueness of $T\bar{T}$ deformed CFT,''
JHEP {\bf 1901}, 086 (2019).
[arXiv:1808.02492 [hep-th]].
}

\lref\PolchinskiRR{
  J.~Polchinski,
  ``String theory. Vol. 2: Superstring theory and beyond.''
}

\lref\AharonyXN{
  O.~Aharony, A.~Giveon and D.~Kutasov,
  ``LSZ in LST,''
Nucl.\ Phys.\ B {\bf 691}, 3 (2004).
[hep-th/0404016].
}

\lref\HullVG{
  C.~M.~Hull,
  ``Timelike T duality, de Sitter space, large N gauge theories and topological field theory,''
JHEP {\bf 9807}, 021 (1998).
[hep-th/9806146].
}

\lref\DijkgraafLYM{
  R.~Dijkgraaf, B.~Heidenreich, P.~Jefferson and C.~Vafa,
  ``Negative Branes, Supergroups and the Signature of Spacetime,''
[arXiv:1603.05665 [hep-th]].
}

\lref\ChakrabortyAJI{
  S.~Chakraborty,
  ``Wilson loop in a $T\bar{T}$ like deformed $\rm{CFT}_2$,''
Nucl.\ Phys.\ B {\bf 938}, 605 (2019).
[arXiv:1809.01915 [hep-th]].
}

\lref\McGoughLOL{
  L.~McGough, M.~Mezei and H.~Verlinde,
  ``Moving the CFT into the bulk with $ T\overline{T} $,''
JHEP {\bf 1804}, 010 (2018).
[arXiv:1611.03470 [hep-th]].
}

\lref\DijkgraafLYM{
  R.~Dijkgraaf, B.~Heidenreich, P.~Jefferson and C.~Vafa,
  ``Negative Branes, Supergroups and the Signature of Spacetime,''
JHEP {\bf 1802}, 050 (2018).
[arXiv:1603.05665 [hep-th]].
}

\lref\KutasovXU{
  D.~Kutasov and N.~Seiberg,
  ``More comments on string theory on AdS(3),''
JHEP {\bf 9904}, 008 (1999).
[hep-th/9903219].
}

\lref\KrausXRN{
  P.~Kraus, J.~Liu and D.~Marolf,
  ``Cutoff AdS$_{3}$ versus the $ T\overline{T} $ deformation,''
JHEP {\bf 1807}, 027 (2018).
[arXiv:1801.02714 [hep-th]].
}

\lref\GuicaNZM{
  M.~Guica and R.~Monten,
  ``$T\bar T$ and the mirage of a bulk cutoff,''
[arXiv:1906.11251 [hep-th]].
}

\lref\HyunJV{
  S.~Hyun,
  ``U duality between three-dimensional and higher dimensional black holes,''
J.\ Korean Phys.\ Soc.\  {\bf 33}, S532 (1998).
[hep-th/9704005].
}

\lref\GiribetIMM{
  G.~Giribet,
  ``$T\bar{T}$-deformations, AdS/CFT and correlation functions,''
JHEP {\bf 1802}, 114 (2018).
[arXiv:1711.02716 [hep-th]].
}

\lref\ForsteWP{
  S.~Forste,
  ``A Truly marginal deformation of SL(2, R) in a null direction,''
Phys.\ Lett.\ B {\bf 338}, 36 (1994).
[hep-th/9407198].
}

\lref\ArgurioTB{
  R.~Argurio, A.~Giveon and A.~Shomer,
  ``Superstrings on AdS(3) and symmetric products,''
JHEP {\bf 0012}, 003 (2000).
[hep-th/0009242].
}

\lref\GiveonMYJ{
  A.~Giveon, N.~Itzhaki and D.~Kutasov,
  ``A solvable irrelevant deformation of AdS$_{3}$/CFT$_{2}$,''
JHEP {\bf 1712}, 155 (2017).
[arXiv:1707.05800 [hep-th]].
}

\lref\GiveonGW{
  A.~Giveon and D.~Gorbonos,
  ``On black fundamental strings,''
JHEP {\bf 0610}, 038 (2006).
[hep-th/0606156].
}

\lref\MaldacenaCG{
  J.~M.~Maldacena and A.~Strominger,
  ``Semiclassical decay of near extremal five-branes,''
JHEP {\bf 9712}, 008 (1997).
[hep-th/9710014].
}

\lref\GiveonGE{
  A.~Giveon, E.~Rabinovici and A.~Sever,
  ``Beyond the singularity of the 2-D charged black hole,''
JHEP {\bf 0307}, 055 (2003).
[hep-th/0305140].
}

\lref\ApoloZAI{
  L.~Apolo, S.~Detournay and W.~Song,
  ``TsT, $T\bar{T}$ and black strings,''
[arXiv:1911.12359 [hep-th]].
}

\lref\CarlipZN{
  S.~Carlip,
  ``Conformal field theory, (2+1)-dimensional gravity, and the BTZ black hole,''
Class.\ Quant.\ Grav.\  {\bf 22}, R85 (2005).
[gr-qc/0503022].
}

\lref\MyersYC{
  R.~C.~Myers,
  ``Myers–Perry black holes,''
[arXiv:1111.1903 [gr-qc]].
}

\lref\AsratUIB{
  M.~Asrat and J.~Kudler-Flam,
  ``$T\bar{T}$, the entanglement wedge cross section, and the breakdown of the split property,''
[arXiv:2005.08972 [hep-th]].
}


\lref\ItzhakiGLF{
  N.~Itzhaki,
  ``Stringy instability inside the black hole,''
JHEP {\bf 1810}, 145 (2018).
[arXiv:1808.02259 [hep-th]].
}

\lref\GiveonXXH{
  A.~Giveon, N.~Itzhaki and U.~Peleg,
  ``Instant Folded Strings and Black Fivebranes,''
[arXiv:2004.06143 [hep-th]].
}

\lref\AttaliGOQ{
  K.~Attali and N.~Itzhaki,
  ``The Averaged Null Energy Condition and the Black Hole Interior in String Theory,''
Nucl.\ Phys.\ B {\bf 943}, 114631 (2019).
[arXiv:1811.12117 [hep-th]].
}



\lref\GiveonDXE{
  A.~Giveon, N.~Itzhaki and D.~Kutasov,
  ``Stringy Horizons II,''
JHEP {\bf 1610}, 157 (2016).
[arXiv:1603.05822 [hep-th]].
}

\lref\ItzhakiRLD{
  N.~Itzhaki and L.~Liram,
  ``A stringy glimpse into the black hole horizon,''
JHEP {\bf 1804}, 018 (2018).
[arXiv:1801.04939 [hep-th]].
}
\lref\GiveonGFK{
  A.~Giveon and N.~Itzhaki,
  ``Stringy Black Hole Interiors,''
JHEP {\bf 1911}, 014 (2019).
[arXiv:1908.05000 [hep-th]].
}
\lref\ItzhakiCGG{
  A.~Giveon and N.~Itzhaki,
  ``Stringy Information and Black Holes,''
[arXiv:1912.06538 [hep-th]].
}

\lref\ChakrabortyNME{
  S.~Chakraborty, A.~Giveon and D.~Kutasov,
  ``Comments on D3-Brane Holography,''
[arXiv:2006.14129 [hep-th]].
}

\lref\BlairOPS{
  C.~D.~A.~Blair,
  ``Non-relativistic duality and $T \bar T$ deformations,''
[arXiv:2002.12413 [hep-th]].
}


\Title{} {\centerline{$T\bar T$, Black Holes and Negative Strings}}

\bigskip
\centerline{\it Soumangsu Chakraborty$^1$, Amit Giveon$^2$ and David Kutasov$^3$}

\bigskip
\smallskip
\centerline{${}^{1}$Department of Theoretical Physics} \centerline{Tata Institute of Fundamental Research, Mumbai 400005, India}
\smallskip
\centerline{${}^{2}$Racah Institute of Physics, The Hebrew
University} \centerline{Jerusalem 91904, Israel}
\smallskip
\centerline{${}^3$EFI and Department of Physics, University of
Chicago} \centerline{5640 S. Ellis Av., Chicago, IL 60637, USA }
\smallskip
\vglue .3cm

\bigskip

\bigskip
\noindent
String theory on $AdS_3$ has a solvable single-trace irrelevant deformation that is closely related to $T\bar T$. For one sign of the coupling, it leads to an asymptotically linear dilaton spacetime, and a corresponding Hagedorn spectrum. For the other, the resulting spacetime has a curvature singularity at a finite radial location, and an upper bound on the energies of states. Beyond the singularity, the signature of spacetime is flipped and there is an asymptotically linear dilaton boundary at infinity. We study the properties of black holes and fundamental strings in this spacetime, and find a sensible picture. The singularity does not give rise to a hard ultraviolet wall for excitations -- one must include the region beyond it to understand the theory. The size of black holes diverges as their energy approaches the upper bound, as does the location of the singularity. Fundamental strings pass smoothly through the singularity, but if their energy is above the upper bound, their trajectories are singular. From the point of view of the boundary at infinity, this background can be thought of as a vacuum of Little String Theory which contains a large number of negative strings.

\bigskip

\Date{}


\newsec{Introduction}

About twenty years ago \GiveonZM, it was pointed out that string theory has a class of weakly coupled backgrounds that interpolate between linear dilaton spacetimes in the ultraviolet (\ie\ at large values of a radial coordinate, where the string coupling goes to zero), and $AdS_3$ in the infrared. Like all asymptotically linear dilaton spacetimes, these backgrounds are holographic \AharonyUB. The natural observables are correlation functions of operators living on the boundary at infinity, which are described in the bulk by non-normalizable wave functions, and physical states, which correspond to normalizable (and delta function normalizable) wave functions. From the point of view of the UV theory, these backgrounds describe states in Little String Theory (LST) \refs{\AharonyUB\AharonyKS-\KutasovUF}, which contain a large number of fundamental strings \refs{\GiveonZM,\GiveonNIE}. This description is somewhat implicit due to the absence of an independent formulation of LST.

After the advent of $T\bar T$ deformed CFT \refs{\SmirnovLQW,\CavagliaODA}, it was noted \GiveonNIE\ that the backgrounds of \GiveonZM\ have many features in common with $T\bar T$ deformed CFT. In particular, from the infrared perspective, these backgrounds correspond to string theory on $AdS_3$, which is dual to a $CFT_2$, in the presence of a deformation which corresponds in the boundary theory to adding to the Lagrangian a certain dimension $(2,2)$ quasi-primary operator. This operator, which  was constructed in \KutasovXU, shares many properties with $T\bar T$, but is distinct from it. Moreover, like $T\bar T$ deformed CFT, the bulk theory is exactly solvable in the presence of this perturbation.\foot{See \eg\ \ChakrabortyMDF\ for further discussion of these and some closely related backgrounds.} Thus, the results of \refs{\GiveonZM,\GiveonNIE,\KutasovXU} provide a constructive definition of the boundary theory holographically related to string theory in a class of asymptotically linear dilaton backgrounds.

The resulting theory is non-local (\eg\ it has a Hagedorn density of states at high energies), and therefore goes beyond the standard QFT paradigm of a UV fixed point connected to an IR fixed point by an RG flow. There was some work on the manifestations of this non-locality in observables such as correlation functions~\refs{\AsratTZD,\GiribetIMM}, spatial entanglement \refs{\ChakrabortyKPR,\AsratUIB} and Wilson lines \ChakrabortyAJI. Some generalizations of the construction were investigated in~\refs{\ChakrabortyMDF,\ChakrabortyVJA\ApoloQPQ-\ApoloZAI}.

An important question concerns the precise relation of this theory to $T\bar T$ deformed CFT. This is an open problem, but one thing that is known about it is the following. String theory on $AdS_3$ has a sector dual to a symmetric product CFT,
which describes long strings~\refs{\ArgurioTB,\GiveonMI}; see \ChakrabortyMDF\ for a recent discussion and further references. In this sector, the deformation of  \refs{\GiveonZM,\GiveonNIE} corresponds to replacing the symmetric product of CFT's by a symmetric product of $T\bar T$ deformed CFT's \refs{\GiveonNIE,\GiveonMYJ,\ChakrabortyMDF}.\foot{For this reason, this deformation is often referred to as a ``single-trace $T\bar T$ deformation.''} This relation was used to obtain new insights into $T\bar T$ deformed CFT and related theories; see \eg\ \refs{\HashimotoWCT,\HashimotoHQO} for recent discussions.

As should be clear from the above comments, in our view it is important to improve the understanding of the backgrounds of \GiveonZM\ and related ones, since they constitute one of the only cases where holography can be studied in detail in a situation where the UV theory is not a CFT.

A central tool in studying the AdS/CFT correspondence has been the analysis of black hole (BH) solutions of the bulk theory. Some preliminary comments on their features in the backgrounds of \GiveonZM\ were made in \GiveonNIE. One of the purposes of this note is to expand on those comments, and apply them to the study of the bulk-boundary duality. In particular, we will show that BH thermodynamics in the background of \GiveonZM\ leads to an energy formula essentially identical to that of $T\bar T$ deformed CFT.

The main focus of this note involves the sign of the coupling in $T\bar T$ deformed CFT. In the original work on this subject \refs{\SmirnovLQW,\CavagliaODA} it was shown that for one sign of the $T\bar T$ coupling, which we will call negative, there is an upper bound on the energies of states in the undeformed theory that give rise to states with real energy after the deformation. This led to questions such as what is the fate of states with energy larger than the bound, and what is the high energy behavior of the theory. Despite some subsequent work, these questions remain open.

There was also some work on the holographic dual of $T\bar T$ deformed CFT with negative coupling, starting with \McGoughLOL. In that paper it was proposed that turning on a negative $T\bar T$ coupling corresponds in the bulk to introducing a UV wall (an upper bound on the radial coordinate) in $AdS_3$. The location of this wall depends on the value of the coupling. As the coupling goes to zero, the wall approaches the boundary of $AdS_3$. The upper bound on the energies corresponds according to this proposal to the constraint that black holes in the resulting background fit in the cut-off $AdS_3$. This proposal was subsequently discussed extensively (see \eg\ \refs{\KrausXRN,\GuicaNZM} and references therein),
but much about it remains mysterious.

The correspondence between the background of \GiveonZM\ and $T\bar T$ deformed CFT holds for positive coupling. One can construct the theory for negative coupling as well, and one of our main goals here will be to study it, in order to shed light on the above questions, and related ones.

We will arrive at a picture which in some ways resembles that of \McGoughLOL. The bulk spacetime approaches $AdS_3$ in the IR, and has a naked singularity in the UV (the analog of the UV wall in $AdS_3$ discussed in \McGoughLOL). Black holes in this background have a maximal energy, at which their horizon approaches the UV singularity.

Our picture also differs in important ways from that of \McGoughLOL. The most significant one is that we will find that the region beyond the (original) UV wall plays an important role in the dynamics. In particular, we will construct BH solutions in the corresponding background, and find that as their energy increases, the radial location of the UV wall (the naked singularity) increases as well, and as the energy approaches the critical energy of \refs{\SmirnovLQW,\CavagliaODA}, the wall is pushed towards the boundary at infinity. We will also study the dynamics of long fundamental strings in this background, and find that they pass smoothly through the singularity, but when their energy exceeds the critical energy of \refs{\SmirnovLQW,\CavagliaODA}, they encounter a (different) singularity beyond it.

Thus, we conclude that in studying the single-trace $T\bar T$ deformed CFT with negative coupling, one needs to include in the description both sides of the naked singularity, and in particular the asymptotically linear dilaton region beyond the singularity.

As mentioned above, single-trace $T\bar T$ deformed CFT with positive coupling can be viewed from the UV point of view as a vacuum of LST with a large number of fundamental strings bound to the fivebranes. We will see that the theory with negative coupling can similarly be viewed as a vacuum of LST with a large number of {\it negative} fundamental strings (see \eg~\DijkgraafLYM\ and references therein for discussions of negative branes in string theory).

The plan of the rest of this note is the following. In section 2, we review the non-extremal background of $k$ Neveu-Schwartz fivebranes, $p$ fundamental strings, and momentum $n$ on the $S^1$ wrapped by the strings. In section 3, we take the near-horizon limit of the fivebranes, and study the resulting BH thermodynamics. In particular, we show that it leads to a formula for energies of states that is closely related to that of $T\bar T$ deformed CFT. In section 4, we study single-trace $T\bar T$ deformed CFT with negative coupling. We show that BH thermodynamics leads in this case to a maximal energy, at which the size of the BH goes to infinity, and its horizon approaches the naked singularity. In section 5, we study the dynamics of probe fundamental strings, and find results which agree with the expected structure in single-trace $T\bar T$ deformed CFT. In section 6, we discuss our results and possible generalizations. Five appendices contain relevant technical results.

\newsec{The black hole background}

The construction of  \GiveonZM\ is quite general, but it will be sufficient for our purposes to consider a special case, type II string theory on $\IR^{4,1}\times T^4\times S^1$,  in the presence of $k$ Neveu-Schwarz fivebranes (NS5-branes) wrapping $T^4\times S^1$ and $p$ fundamental strings wrapping the $S^1$, with $n$ units of momentum on the $S^1$.
The metric, dilaton and NS three-form field strength take in this background the form~\refs{\MaldacenaKY,\HyunJV}\foot{This background can be obtained \eg\ by starting with that of $k$ near-extremal NS5-branes~\MaldacenaCG, and performing a boost followed by a T-duality and another boost (see \eg~\GiveonGW, for a description of the procedure), thus turning on fundamental string winding $p$ and momentum $n$, given in terms of the boost parameters, $\alpha_{1,n}$, and the other parameters below.}
\eqn\background{\eqalign{ ds^2=&\frac{1}{f_1}\left[-\frac{f}{f_n}dt^2+f_n\left(dx+\frac{r_0^2\sinh 2\alpha_n}{2f_n r^2}dt\right)^2\right]+f_5\left(\frac{1}{f}dr^2+r^2 d\Omega_3^2\right)+\sum_{i=1}^4dx_i^2,\cr
 e^{2\Phi} = &g^2\frac{f_5}{f_1}, \cr
 H=& dx \wedge dt \wedge d\left(\frac{r_0^2\sinh2\alpha_1}{2f_1r^2}\right)+r_0^2\sinh2\alpha_5d\Omega_3,}}
where $g=e^{\Phi_0}$ is the string coupling far from the branes, related to the ten dimensional Newton constant in flat spacetime by
\eqn\gnten{G_{N}^{(10)}=8\pi^6g^2l_s^8\;,}
and $l_s=\sqrt{\alpha'}$ is the string length.  The coordinate $x$ parametrizes a circle of radius $R$, $x_i$ $(i=1,\cdots,4)$, label the $T^4$, while $r$ and $\Omega_3$ are spherical coordinates on the $\IR^4$ transverse to the branes.

The harmonic functions in \background\ are given by
\eqn\harfun{f=1-\frac{r_0^2}{r^2}, \ \ \ \ f_{1,5,n}=1+\frac{r_{1,5,n}^2}{r^2}, \ \ \ \ r^2_{1,5,n}=r_0^2\sinh^2\alpha_{1,5,n}.}
The relations between the parameters $\alpha_{1,5,n}$  and the integer charges $p,k,n$ are given by
\eqn\charges{\sinh2\alpha_1=\frac{2l_s^2p}{vX}, \ \ \ \ \sinh2\alpha_5=\frac{2l_s^2k}{r_0^2}, \ \ \ \  \sinh2\alpha_n=\frac{2l_s^4n}{R^2vX},}
with
\eqn\X{X=\frac{r_0^2}{g^2}.}
The dimensionless quantity $v$ is related to the volume of the $T^4$ by\foot{Following the conventions of~\MaldacenaKY.}
\eqn\vtfour{V_{T^4}=(2\pi)^4vl_s^4.}
The parameter $r_0$ above is a non-extremality parameter. Taking it to zero, and setting the momentum on the $S^1$, $n$, to zero, gives the supersymmetric geometry describing $p$ fundamental strings and $k$ NS5-branes in asymptotically flat spacetime.

For $r_0>0$, \background\ describes a black brane in flat spacetime. It has an outer horizon at $r=r_0$ and an inner one at $r=0$. As usual, the thermodynamics of the outer horizon can be used to learn about the properties of highly excited states with the charges $(p,k,n)$ in string theory.

The ADM mass of the non-extremal geometry \background\ is
\eqn\mADM{M_{\rm ADM}=\frac{R vX}{2l_s^4}(\cosh2\alpha_1+\cosh2\alpha_5+\cosh2\alpha_n)}
or, equivalently, in terms of the number of strings $(p)$, fivebranes $(k)$, and momentum $(n)$,
\eqn\mkpn{M_{\rm ADM}=\frac{R}{l_s^2}\sqrt{p^2+\frac{X^2v^2}{4l_s^4}}+\frac{R v}{g^2l_s^2}\sqrt{k^2+\frac{X^2g^4}{4l_s^4}}+\frac{1}{R}\sqrt{n^2+\frac{X^2R^4v^2}{4l_s^8}}.}
The extremal energy, which corresponds to $X=n=0$ in \mkpn, is given by
\eqn\plpl{E_{\rm ext}=p{R\over l_s^2}+k{\frac{R v}{g^2l_s^2}}=\frac{R vX}{2l_s^4}(\sinh2\alpha_1+\sinh2\alpha_5),
}
from which we can read off the tension of a fundamental string, $T_1=1/2\pi\alpha'$, and NS5-brane, $T_5=1/g^2(2\pi)^5l_s^6$, which agree with the standard conventions of~\PolchinskiRR.

The energy of the black hole \background\ above extremality is given by
\eqn\eaext{E=M_{\rm ADM}-E_{\rm ext}=\frac{R vX}{2l_s^4}(e^{-2\alpha_1}+e^{-2\alpha_5}+\cosh2\alpha_n).}
The entropy is given by \MaldacenaKY
\eqn\ent{S=\frac{2\pi R v X r_0}{l_s^4}\sqrt{f_1f_5f_n}=\frac{2\pi R v X r_0}{l_s^4}\cosh\alpha_1\cosh\alpha_5\cosh\alpha_n,}
where all the quantities are evaluated at the outer horizon, $r=r_0$. The inverse Hawking temperature is given by
\eqn\temp{\beta=2\pi r_0 \cosh\alpha_1\cosh\alpha_5\cosh\alpha_n.}
Equation \ent\ is the Bekenstein-Hawking entropy $S=A/4G_N$,
\eqn\sten{S=\frac{2\pi R\sqrt{\frac{f_n}{f_1} }V_{S^3} V_{T^4}}{4G_N^{(10)}e^{2(\Phi-\Phi_0)}},}
where the numerator is the area of the horizon, the eight dimensional surface $S^1\times S^3\times T^4$, evaluated with the metric \background\ at $r=r_0$. In particular, the circumference of the $x$ circle is $2\pi R \sqrt{f_n\over f_1}$, $V_{S^3}=2\pi^2(f_5r_0^2)^{3\over2}$, and $V_{T^4}$ is given by \vtfour. The denominator in \sten\ is $4G_N$ evaluated on the horizon. From the three dimensional perspective used in \GiveonNIE\ one has
\eqn\sthree{S=\frac{2\pi R}{4G_{N}^{(3)}}\frac{\sqrt{f_1f_n}}{f_5},}
where $G_N^{(3)}$ is given by
\eqn\gnr{G_{N}^{(3)}=\frac{G_{N}^{(10)}}{V_{S^3}V_{T^4}}.}
It is related to the Newton constant at the horizon of the black hole, $G_3$, via the relation
\eqn\correctg{G_3=G_{N}^{(3)}e^{2(\Phi-\Phi_0)}=\frac{g^2l_s^4}{4vr_0^3\sqrt{f_5}f_1}}
with $f_{1,5}$ evaluated at the horizon, $r=r_0$, as before.

\newsec{The decoupling limit and BH thermodynamics on $\MM_3$}

As is familiar from discussions of LST (see \eg\ \refs{\AharonyUB\AharonyKS-\KutasovUF,\AharonyXN} and references therein), we can obtain a decoupled theory of fivebranes by taking the asymptotic string coupling $g\to 0$, and focusing on radial distances $r$ of order $gl_s$.  In particular, in this limit the quantity $X$ \X\ is held fixed (in string units). Looking back at \charges, we see that in this limit $\alpha_5\to\infty$,
\eqn\alphak{e^{2\alpha_5}\simeq\frac{4l_s^2k}{r_0^2},}
and
\eqn\ffive{f_5(r)\simeq\frac{kl_s^2}{r^2}\;,}
which is equivalent to dropping the 1 from the harmonic function $f_5$ in the background \background, \ie\ focusing on the near-horizon region of the fivebranes. In this limit, the contribution of the fivebranes to the ADM mass \mkpn\ is equal to the extremal one \plpl. Equivalently, their contribution to the energy above extremality, \eaext, vanishes.

On the other hand, the quantity $\alpha_1$ in \charges\ remains finite in the decoupling limit, and varies between zero and infinity. In the extremal case $r_0=n=0$, the resulting geometry \background\  is that of \GiveonZM, $\MM_3\times S^3\times T^4$, while for $r_0>0$ it describes a (rotating) charged black string in $\MM_3$ times $S^3\times T^4$. The charge of the black string is $p$, while its angular momentum is $n$.

To study the thermodynamics of the non-extremal background \background\ in the decoupling limit, we consider the special case $n=0$. The background takes in this case the form
\eqn\nzero{\eqalign{ ds^2=&\frac{1}{f_1}\left(-fdt^2+dx^2\right)+f_5\left(\frac{1}{f}dr^2+r^2 d\Omega_3^2\right)+\sum_{i=1}^4dx_i^2,\cr
 e^{2\Phi} = &g^2\frac{f_5}{f_1}, \cr
 H=& dx \wedge dt \wedge d\left(\frac{r_0^2\sinh2\alpha_1}{2f_1r^2}\right)+2kl_s^2d\Omega_3}}
with $f_1$ given by \harfun, and $f_5$ given by \ffive.

Setting $n=0$ in \charges, \eaext\ and \ent\ and taking the decoupling limit, one finds the entropy
\eqn\entnzero{S=\frac{\pi R r_0^2\cosh\alpha_1}{2G_{N}^{(3)}kl_s^2}=2\pi\sqrt{\frac{kl_s^2{\cal{E}}^2}{R^2}+2kp{\cal{E}}},}
where
\eqn\eee{{\cal{E}}=ER.}
Thus, for low energies the entropy goes as $2\pi\sqrt{2kpRE}$, the Cardy entropy for a CFT with central charge $c=6kp$
and scaling dimensions $h+\bar h-c/12=ER$, $h-\bar h=0$ (see appendix A).
For high energies, the entropy goes as $\beta_h E$, where $\beta_h$ is the inverse Hagedorn temperature given by
\eqn\hag{\beta_h=2\pi\sqrt{k}l_s.}

Using \harfun, \charges\ and \eaext\ one can write
\eqn\ronerzero{\frac{r_1^2}{r_0^2}=\frac{{\cal{E}}_c^2}{({\cal{E}}+{\cal{E}}_c)^2-{\cal{E}}_c^2},}
where
\eqn\ec{{\cal{E}}_c=\frac{pR^2}{l_s^2}}
is the energy at which the system transitions between the Cardy and Hagedorn regimes \entnzero. As explained in \refs{\GiveonNIE,\GiveonMYJ},
one can think of this (dimensionless, \eee) energy as $1/\lambda$, where $\lambda=l_s^2/R^2$ is the analog of the $T\bar T$ coupling evaluated at the scale $R$. The factor of $p$ in \ec\ has to do with the fact that in the analogy to $T\bar T$ deformed CFT, the CFT that is being deformed is the block of a symmetric product, so the natural object to consider is $\EE/p$, the energy per block \GiveonNIE.

For $\EE\ll\EE_c$, we have $r_1^2/r_0^2\simeq \EE_c/2\EE\gg 1$, so the relevant harmonic function in \harfun, $f_1$ (evaluated at the horizon), behaves as $f_1\sim r_1^2/r_0^2$, and the horizon of the BH is deep inside the $AdS_3$ region in $\MM_3$. On the other hand, for $\EE\gg\EE_c$, $r^2_1/r^2_0\simeq \EE_c/\EE\ll 1$, $f_1\simeq 1$, and the horizon is deep inside the linear dilaton  region in $\MM_3$.  For general $\EE$ one finds, using \charges, \eaext,  \ronerzero,\foot{Here we use the original variables in terms of which \nzero\ is written. In the decoupling limit, it is more useful to use the parametrization introduced later, in eq. (5.1), in terms of which the limit $g\to 0$, $p\to\infty$ is smooth.}
\eqn\rone{\eqalign{r_1^2&=4pk^{3/2}l_sG_N^{(3)}\left(\frac{{\cal{E}}_c}{{\cal{E}}+{\cal{E}}_c}\right),\cr
r_0^2&=4pk^{3/2}l_sG_N^{(3)}\frac{{\cal{E}}({\cal{E}}+2{\cal{E}}_c)}{{\cal{E}}_c({\cal{E}}+{\cal{E}}_c)}.}}

\smallskip

\noindent
We can use the above discussion to derive an energy formula which is very similar to the one familiar from $T\bar T$ deformed CFT \refs{\SmirnovLQW,\CavagliaODA}. The basic idea is the following.

We can view the dimensionless combination
\eqn\lamb{\lambda=\frac{l_s^2}{R^2}}
 as a coupling parametrizing a line of theories.  As we vary this parameter, the energies of states change. We can follow the way they change, by noting that any two states that are degenerate at a particular value of the coupling remain degenerate for all values of the coupling. This is certainly a feature of (single) perturbative long string states in $\MM_3$, and of $T\bar T$ deformed CFT.\foot{ For example,  it played an important role in \AharonyBAD.} We will assume that it holds for generic high energy states, that are represented by black holes in $\MM_3$.

Thus, consider a BH with energy $\EE$ at a particular value of the coupling $\lambda$. This BH provides a thermodynamic description of a large class of microscopic states with that energy. The entropy of the BH is the log of the number of these states. As we vary $\lambda$, the energy of these states changes, but since it changes in the same way for all $e^S$ states (due to the above universality), the number of such states, and thus their entropy, remains the same. Therefore, if we want to follow the way the energy of a given state changes with the coupling, we can use the entropy of states at that energy as a coupling-independent characterization of the state.

Thus, we can write \entnzero\ as\foot{Note that since we are using classical gravity in the present discussion, strictly speaking it is only applicable in the limit $1\ll k\ll p$. In particular, for classical black holes the energy and entropy go like $p$. Our results are expected to receive $1/p$ corrections, which we neglect.}
\eqn\entnone{S=2\pi\sqrt{k\lambda\EE(\lambda)^2+2kp\EE(\lambda)},}
where the l.h.s. does not depend on $\lambda$, so we can evaluate it at any value of $\lambda$ we choose, say $\lambda=0$. We find that $S=2\pi\sqrt{2kp\EE(0)}$,~\foot{This is the entropy of a BTZ black hole with mass $M=\EE(0)/\sqrt{k}l_s$, and angular momentum $n=0$;
see appendices A and B for a short review.}
with $\EE(0)$ being the energies of the states in the undeformed CFT that become after the deformation states with energy $\EE(\lambda)$.

Plugging the expression for the entropy into \entnone\ leads to a quadratic equation for the deformed energies, whose solution is
\eqn\solspec{{1\over p}\EE(\lambda)={1\over\lambda}\left(-1+\sqrt{1+2\lambda\EE(0)/p}\right).
}
This is precisely the spectrum found in \refs{\SmirnovLQW,\CavagliaODA}, written in terms of $\EE/p$, the energy per block in a symmetric product of $p$ $T\bar T$ deformed CFT's. This agreement suggests that generic high energy states in $\MM_3$ are well described by such a symmetric product.

The discussion above was for states with zero momentum along the $S^1$, $n=0$. The generalization to the case of arbitrary $n$ is presented in appendix B.

\newsec{Negative coupling}

The deformation that takes string theory on $AdS_3$ to $\MM_3$ is a Thirring-type perturbation of the worldsheet theory \refs{\ForsteWP,\GiveonNIE}. It correspond to adding to the Lagrangian of the worldsheet theory a term
\eqn\deltal{\delta\CL_{\rm ws}=\mu J^-\bar J^-,}
where $J^-$ and $\bar J^-$ are particular left and right-moving worldsheet currents. The construction in the previous sections involves turning on the deformation \deltal\ with a specific sign.

From the worldsheet point of view, nothing prevents us from turning on the perturbation \deltal\ with the opposite sign. In fact, this deformation preserves $(4,4)$ spacetime supersymmetry for both signs of the coupling. However, the resulting string backgrounds are quite different. The background $\MM_3$ studied in the previous section is
\eqn\mthree{ds^2=\frac{1}{f_1}\left(-dt^2+dx^2\right)+kl_s^2\frac{dr^2}{r^2}~,\qquad
e^{2(\Phi-\Phi_0)}=\frac{kl_s^2}{r^2f_1}~ ,\qquad  B_{tx}= \frac{1}{f_1}~,}
with the harmonic function $f_1$  given by
\eqn\fone{f_1=1+\frac{r_1^2}{r^2}~,\qquad r_1^2={g^2l_s^2p\over v}.}
It interpolates between $AdS_3$ (or, more precisely, massless BTZ; see appendix A) for small $r$, and a linear dilaton spacetime $R_t\times S^1\times R_\phi$ for large $r$.

The other sign of $\mu$ in \deltal\ gives rise to a spacetime of the form \mthree, but in this case the harmonic function $f_1$ assumes the form  \GiveonNIE
\eqn\fonet{ f_1=-1+\frac{r_1^2}{r^2}~,}
with $r_1$ the same as before, \fone. Comparing \fonet\ to \fone, we see that the behavior at small $r$, which corresponds to the infrared limit of the boundary theory, is again massless BTZ, but as we increase $r$ we find a qualitatively different background. In particular, in the case \fonet, as $r\to r_1$ we approach a curvature singularity which is not shielded by a horizon (see figure 1(a)).  The dilaton in \mthree\ diverges at the singularity as well.

\ifig\loc{(a) $\MM_3^-$ interpolates between a massless BTZ black hole at $r=0$ (denoted in yellow) and a naked singularity at $r=r_1$ (dashed red line). The region beyond the singularity, $r>r_1$, must be included; it is asymptotically a flat spacetime with a linear dilaton and flipped signature of $(t,x)$.  From the point of view of an observer at infinity, this background is formed by adding $p$ negative strings to the near-horizon geometry of $k$ NS5-branes. (b) A black hole with a finite energy in $\MM_3^-$ has a singularity at $r=0$ and horizon at $r=r_0$. The location $r_1$ of the naked singularity increases when $r_0$ increases, and is always outside the black hole $(r_1>r_0)$. (c) As the energy of the black hole approaches its maximal value, its horizon goes to infinity
and approaches $r_1$ asymptotically.}
{\epsfxsize4.3in\epsfbox{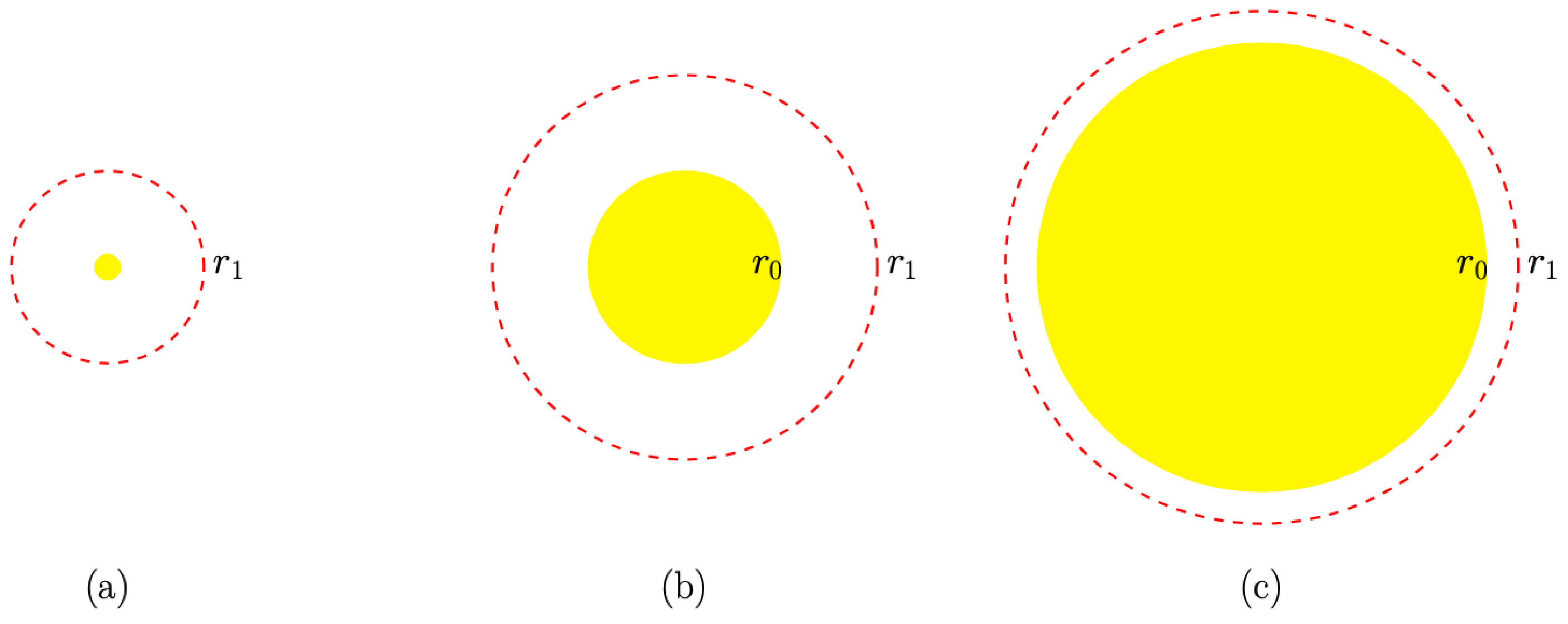}}

If we continue past the singularity, to $r>r_1$, $f_1$ flips sign, $t$ becomes a spacelike coordinate and $x$ becomes timelike. Superficially, one might think that the regions on the two sides of the singularity at $r=r_1$ should be considered separately, but we will argue below that this is not the case.\foot{Arguments that one should include both sides of related singularities have a long history; see \eg\ \GiveonFU\ for a review, and \GiveonGE\ and references therein, for a more recent discussion.} Moreover, we will argue that for the purpose of studying single-trace $T\bar T$ deformed CFT with negative coupling, it is advantageous to view $t$ as time and $x$ as space, despite the fact that near the boundary at large $r$ they are spacelike and timelike, respectively.


We will refer to the full background \mthree, \fonet\ as $\MM_3^-$. From this point of view, the background \mthree, \fone\ can be referred to as $\MM_3^+$. The backgrounds $\MM_3^+$ and $\MM_3^-$ are formally related by $r^2\to -r^2$, $t\leftrightarrow x$.
However, it is important to stress that they are not geodesically connected~\HorneGN.
This is in agreement with the fact that they are different theories,
corresponding to the (non-normalizable) perturbation \deltal\ with opposite signs of $\mu$.

Our goal is to repeat the analysis of section 3 for $\MM_3^-$. Thus, we would like to construct a BH, whose horizon $r_0$ is located in the region $0<r_0<r_1$. One can check that the following metric, dilaton and $H$ field satisfy the supergravity equations of motion (see also \ApoloZAI):
\eqn\mtmft{\eqalign{ ds^2=&\frac{1}{f_1}\left(-fdt^2+dx^2\right)+f_5\left(\frac{1}{f}dr^2+r^2 d\Omega_3^2\right)+\sum_{i=1}^4dx_i^2~,\cr
 e^{2\Phi} = &g^2\frac{f_5}{f_1}~, \cr
 H=& dt \wedge dx \wedge d\left(\frac{r_0^2\sinh2\alpha_1}{2f_1r^2}\right)+2kl_s^2d\Omega_3~,}}
where $f$, $f_1$ and $f_5$ are given by\foot{Here we already took the decoupling limit of section 3. The background \mtmft\ is a solution of the equations of motion also when $f_5$ is taken to have the form \harfun, \charges, \ie\ before taking the decoupling limit (with the H-flux replaced by that in \background).}
\eqn\harf{f=1-\frac{r_0^2}{r^2}~, \ \ \ f_1=-1+\frac{r_1^2}{r^2}~, \ \ \ f_5=\frac{kl_s^2}{r^2}~,}
and
\eqn\rrcc{r_1^2=r_0^2\cosh^2\alpha_1~.}
The relation between $\alpha_1$ and $p$ is given by the first equation in \charges.

Other than the UV singularity at $r=r_1$, the background \mtmft\ -- \rrcc\ looks like a standard BH spacetime, with a horizon at $r=r_0$ and a singularity at $r=0$.  The Bekenstein-Hawking entropy associated with the horizon is given by
\eqn\entm{S=\frac{2\pi R}{4G_{N}^{(3)}}\frac{\sqrt{f_1}}{f_5}=\frac{\pi R r_0^2\sinh\alpha_1}{2G_{N}^{(3)}kl_s^2},}
where all the quantities are evaluated at $r=r_0$. Wick rotating to Euclidean time, $t\to i\tau$, compactifying $\tau$ on a circle of circumference $\beta=1/T$, and demanding regularity of the metric \mtmft\ at the horizon, one finds the inverse Hawking temperature
\eqn\hawt{\beta=2\pi \sqrt{k} l_s \sinh\alpha_1.}
We can use BH thermodynamics to compute the entropy-energy relation, $S=S(E)$, by using the fact that $E=E(r_0)$, $E(0)=0$, and imposing the thermodynamic relation
\eqn\thermod{\frac{\partial S}{\partial E}=\beta.}
This gives
\eqn\entme{S=2\pi\sqrt{-\frac{kl_s^2{\cal{E}}^2}{R^2}+2kp{\cal{E}}}={2\pi R\over l_s}p\sqrt{k}\sqrt{\EE_c^2-(\EE_c-\EE)^2\over\EE_c^2}~,}
where ${\cal{E}}=ER$, ${\cal{E}}_c=pR^2/l_s^2$, as before. $E(r_0)$ is given by
\eqn\eaextm{E=\frac{R vX}{2l_s^4}(-e^{-2\alpha_1}+1).}
The entropy \entme\ vanishes for $\EE=0$, and reaches its maximal value for $\EE=\EE_c$. To see what that means geometrically, it is convenient to write \entme\ in terms of $r_0$, $r_1$. Using \charges\ and \eaextm, one finds\foot{Compare to \ronerzero, which is the analogous relation for positive coupling.}
\eqn\ronerzerom{\frac{r_1^2}{r_0^2}=\frac{{\cal{E}}_c^2}{{\cal{E}}_c^2-({\cal{E}}_c-{\cal{E}})^2}~,}
from which we see that $r_0\le r_1$ for all $\EE$. The entropy \entme\ can be written using \ronerzerom\ as
\eqn\entmenew{S={2\pi R\over l_s}p\sqrt{k}{r_0\over r_1}~.}
The maximal entropy case corresponds to $\EE\to\EE_c$, or $r_0\to r_1$.

It is instructive to generalize the calculation of $r_0$, $r_1$ in section 3 (eq. \rone) to the case of negative coupling. One finds
\eqn\ronem{\eqalign{r_1^2&=4pk^{3/2}l_sG_N^{(3)}\left(\frac{{\cal{E}}_c}{{\cal{E}}_c-{\cal{E}}}\right),\cr
r_0^2&=4pk^{3/2}l_sG_N^{(3)}\frac{{\cal{E}}(2{\cal{E}}_c-{\cal{E}})}{{\cal{E}}_c({\cal{E}}_c-{\cal{E}})}.}}
As $\EE$ varies between $0$ and $\EE_c$, $r_0$ varies between $0$ and infinity, and $r_1$ varies between the finite value \fone\ and infinity
(see figure 1). The behavior of $r_0$ is interesting since usually the size of a BH goes to infinity in the limit of infinite energy, while here it happens for finite energy. The entropy also has a finite limit as $r_0\to r_1$, \entmenew. Note that since $r_0, r_1\to\infty$ as $\EE\to\EE_c$, $\EE_c$ is a limiting energy in this system -- it doesn't make sense to talk about BH's with energy larger than $\EE_c$.

The above discussion shows that to the extent that the naked singularity at $r=r_1$ provides a UV wall, it is a soft one -- its location changes with the energy we put in the small $r$ region. In particular, we cannot restrict attention to the region $r<r_1$, where $r_1$ is the location of the UV wall in the vacuum, given in eq. \fone, since BH's of energy approaching the critical one have a horizon located at parametrically larger $r$.

We can also ask what happens to the temperature of the BH \hawt\ as we vary the energy. Plugging \rrcc\ into \hawt, we find
\eqn\bebe{\beta=2\pi \sqrt{k} l_s \sqrt{\left(r_1\over r_0\right)^2-1}={2\pi \sqrt{k} l_s(\EE_c-\EE)\over\sqrt{\EE_c^2-(\EE_c-\EE)^2}}~.}
The temperature goes to zero as $\EE\to 0$, and diverges as $\EE\to\EE_c$. The specific heat of these BH's is positive for all $\EE<\EE_c$.

We finish this section with a few comments:
\item{(1)} In section 3, we derived the energy formula \solspec\ from the entropy-energy relation \entnone. It is straightforward to repeat the analysis for the case where this relation is given by \entme, instead. In fact, \entme\ is identical to \entnone, with the coupling \lamb\ replaced by
\eqn\lambm{\lambda=-\frac{l_s^2}{R^2}~.}
Hence, one finds again the spectrum \solspec, but with the negative coupling \lambm.
\item{(2)} The authors of \McGoughLOL\ proposed a holographic dual to $T\bar T$ deformed CFT with negative coupling. The bulk spacetime in their construction is $AdS_3$ with a UV wall at a finite value of the radial coordinate. This value depends on the coupling constant, and goes to infinity (\ie\ to the boundary of $AdS_3$) as the coupling goes to zero. High energy states are BTZ BH's whose horizon is constrained to lie inside the cut-off $AdS_3$. The maximal energy states correspond to a BH whose horizon coincides with the UV wall.  It is instructive to compare our construction to theirs. Of course, the two constructions describe different theories (single and double-trace $T\bar T$ deformations of the boundary CFT, respectively), but it is natural to compare their qualitative features. Some things do agree between the two approaches. The analog of the UV wall of \McGoughLOL\ in our construction is the singularity at $r=r_1$, with $r_1$ given in \fone.  We also find a maximal energy, which corresponds to a BH whose horizon approaches the UV wall. And, the dependence of the energy on the coupling gives in both constructions the result expected from the relevant $T\bar T$ deformed CFT. The main qualitative difference between our construction and that of \McGoughLOL\ is that in our construction, as the energy of the BH increases, so does the position of the UV wall, $r_1$ (see \ronem). The horizon only coincides with the UV wall in the limit $r_1\to\infty$ (see figure 1). Thus, questions like what happens to BH's whose horizon does not fit in the cutoff $AdS_3$ do not arise in our case.
\item{(3)} In studying the thermodynamics of the BH \mtmft\ -- \rrcc, we focused on the region near the horizon of the BH, $r=r_0$. For standard black holes, one can alternatively perform an analysis at infinity. For example, the dependence of the energy of the BH on its size, $E(r_0)$ \eaextm, is usually obtained from the ADM construction \MyersYC, by studying the metric at large $r$. In our case, this asymptotic analysis is much more subtle, since for $r>r_1$ \ronem, the signature of $x$ and $t$ flips sign, as does $\exp(2\Phi)$ \mtmft. In appendix C we describe an ADM-type calculation that has the property that it reproduces the near-horizon analysis of this section. An important feature of that analysis is that we still need to treat $t$ as the time coordinate, despite the fact that at large $r$ it is spacelike. As we explain in appendix C, this is physically reasonable, since we do something similar when we Wick rotate QFT's from Lorentzian to Euclidean signature. The analysis of appendix C gives the ADM mass
\eqn\mADMnew{M_{\rm ADM}=\frac{R vX}{2l_s^4}(-\cosh2\alpha_1+\cosh2\alpha_5+1).}
In the limit $r_0\to 0$, \mADMnew\ takes the form
\eqn\plplpl{E_{\rm ext}=-p{R\over l_s^2}+k{\frac{R v}{g^2l_s^2}}=\frac{R vX}{2l_s^4}(-\sinh2\alpha_1+\sinh2\alpha_5),
}
which looks like the energy of a BPS system of $k$ NS5-branes and $p$ negative strings.
In the decoupling limit of section 3, the energy $E=M_{\rm ADM}-E_{\rm ext}$ takes the form \eaextm.
\item{(4)} The discussion above was for states with zero momentum along the $S^1$, $n=0$. It can be generalized to the case of arbitrary $n$.

\newsec{String probe dynamics in $\MM_3$}

As mentioned in the introduction, the dynamics of long strings in $\MM_3^+$ $(\MM_3^-)$ is described by a symmetric product of $T\bar T$ deformed CFT's with positive (negative) coupling. For negative coupling, in each block of the symmetric product the spectrum of energies is bounded from above, $\EE_{\rm block}<\EE_c/p$ or, equivalently, $E_{\rm block}<R/l_s^2$. In the last section, we saw that the bound on the total energy, $\EE_{\rm total}<\EE_c$, can be understood from the bulk point of view by studying black holes in $\MM_3^-$. In this section, we will try to extend this understanding to an individual block.

To study a perturbation of a single block of the symmetric product, we will discuss the dynamics of a probe long fundamental string in $\MM_3^-$. We can think of this string either as one of the $p$ strings that are used in the construction of the undeformed $AdS_3\times S^3\times T^4$, or as an additional string we add to the system. The difference between the two pictures is only in the value of $p$, which we take to be large throughout this note.

As mentioned above, the duality predicts that the energy of such a string is bounded from above by $R/l_s^2$.  Our goal will be to see if that is indeed the case. As a warm-up exercise, we will first consider the dynamics of such a string in $\MM_3^+$, where we do not expect any bound on the energy, and then turn to the case of interest, $\MM_3^-$.

\subsec{Probe string in $\MM_3^+$}

The starting point of our analysis is the background \mthree, \fone. We parametrize the radial coordinate in terms of $\phi$, related to $r$ via the relation
\eqn\rrpp{r=r_1e^{\frac{\phi}{\sqrt{k}l_s}}.}
Note that $r_1$ in \rrpp\ is given by \fone, and should not be confused with \rone. The precise relation is: $r_1^{(\rm here)}=r_1^{(4.3)}=r_1^{(3.9)}(\EE=0)$.

In the parametrization \rrpp, the harmonic function $f_1$, \fone, takes the form
\eqn\harrphi{f_1=1+e^{-{2\phi\over \sqrt{k}l_s}}.}
In this background, we study a probe fundamental string wrapping the $x$ circle, and located at the point $\phi(t)$ in the radial direction.  The dynamics of such a string is described by the Lagrangian
\eqn\laggg{{\cal L}=-\frac{1}{2\pi l_s^2f_1}\left(\sqrt{1-f_1\dot{\phi}^2}- f_1B_{tx}\right).}
The first term in the brackets is the Nambu-Goto Lagrangian, while the second is the standard coupling of a fundamental string to the $NS$ B-field \PolchinskiRR. As a simple check of the relative coefficient between the two terms, setting $\dot\phi=0$, we find that the potential for the string vanishes, in agreement with the fact that it is a BPS object.

The canonical momentum of $\phi$ is given by
\eqn\cannn{\Pi=\frac{\delta {\cal L}}{\delta \dot{\phi}}=\frac{1}{2\pi l_s^2} \frac{\dot{\phi}}{\sqrt{1-f_1\dot{\phi}^2}}~,}
and the energy of the string is
\eqn\energyyy{E=\int_0^{2\pi R} dx( \Pi \dot{\phi}-{\cal L})=\frac{R}{l_s^2f_1}\left(\frac{1}{\sqrt{1-f_1\dot{\phi}^2}}-f_1B_{tx}\right).}
Plugging in the expressions for $f_1$, \harrphi, and $B_{tx}=1/f_1$, we see that for a stationary string $(\dot\phi=0)$ the energy vanishes,  as appropriate for a BPS object, and for any non-zero velocity the energy is positive.

We can use eq. \energyyy\ to solve for the velocity of the string:
\eqn\phidottt{\dot{\phi}^2=\frac{l_s^2E(2R+l_s^2f_1E)}{(R+l_s^2f_1E)^2}.}
As an example of the resulting motion, we can consider a string with a small energy  $(E\ll R/l_s^2)$, that starts deep in the $AdS_3$ region at $t=0$, $\phi(t=0)=\phi_0$, with $\phi_0$ large and negative $(f_1(\phi_0)E\gg R/l_s^2)$, moving in the positive $\phi$ direction. The initial velocity of such a string, $\dot\phi\simeq 1/\sqrt{f_1}$, is very small for large $|\phi_0|$. As it moves towards larger $\phi$ it accelerates, and for large positive $\phi$, in the region where $f_1(\phi)\simeq 1$, its velocity approaches a constant value, $\dot\phi\simeq l_s\sqrt{2E/R}$. It takes the string an infinite amount of time to reach the boundary, which is located at $\phi\to\infty$.

It is not essential, but we mention for completeness that one can in fact solve the equation of motion \phidottt\ exactly for all $E$. The solution takes the form
 \eqn\ssss{F(\phi(t))-F(\phi_0)=t,}
where
\eqn\FFF{F(\phi(t))= \sqrt{\frac{k}{E}}\left(-\sqrt{2R+l_s^2f_1E}+\frac{R+l_s^2E}{\sqrt{2R+l_s^2E}}{\rm arctanh}\sqrt{\frac{2R+l_s^2E}{2R+l_s^2f_1E}}\right).}
The solution \ssss, \FFF, describes a string moving to the right. The solution for a string moving to the left has a minus sign in front of the r.h.s. in \ssss.

The classical string solutions described above give rise in the quantum theory to a continuum of long string states labeled by the energy $E$. This energy takes values in $\IR_+$, and in particular there is no upper bound on it. We will next see that the situation is different for long strings in $\MM_3^-$.

\subsec{Probe string in $\MM_3^-$}

To study a probe fundamental string in $\MM_3^-$, we need to repeat the analysis of the previous subsection in the background \mthree, \fonet. The only difference between the two is in the form of $f_1$, \fone\ versus \fonet. Thus, we can use the results obtained there, with the appropriate form of $f_1$.

To find the trajectory of the string, we need  to solve eq. \phidottt\ with $f_1$ given by \fonet, \rrpp. In the analogous calculation in $\MM_3^+$, discussed in the previous subsection, the trajectory was regular since the function $f_1(\phi)$, \harrphi, is bounded from below, $f_1>1$,
for all $\phi$. In $\MM_3^-$, $f_1=-1+e^{-{2\phi\over \sqrt{k}l_s}}$ is positive for large negative $\phi$, but it goes to zero at the singularity, and then becomes negative, approaching $-1$ at large positive $\phi$. This leads to a difference in the resulting trajectories.

One important qualitative fact is that the trajectory of the string is smooth at the singularity $r=r_1$, where $f_1$ vanishes. Indeed, the velocity $\dot\phi$ and all higher time derivatives of $\phi$ are finite at that point. This is compatible with the expectation from section 4, that the singularity does not provide a hard wall for excitations of $\MM_3^-$.\foot{Different arguments
for the smoothness of related singularities were presented \eg\ in~\GiveonGE\ and references therein.}

Continuing the trajectory of the string past the singularity, we see that there are two distinct regimes of energy. For $E<R/l_s^2$, the solution remains smooth for all $t$. At late time, $\phi\to\infty$, so $f_1\to -1$, and the velocity \phidottt\ approaches a constant value
\eqn\phiass{\dot{\phi}={l_s\sqrt{E(2R-l_s^2E)}\over  (R-l_s^2E)}~.}
Of course, in that region $t$ is a spacelike coordinate, so it's not clear that one should view $\dot\phi$ as a velocity. We will return to this point later.

For $E>R/l_s^2$, the e.o.m. \phidottt\ is singular at a finite value of $\phi$,
\eqn\phistar{\phi^*=\half \sqrt{k}l_s\log\left(El_s^2\over  El_s^2-R\right).
}
Near the singularity, it takes the form
\eqn\nearsig{\dot\phi\simeq-{\alpha\over\phi-\phi^*}~;\qquad \alpha={l_s^2\sqrt{kER/ 4}\over El_s^2-R}~,
}
with the solution
\eqn\solphi{\phi-\phi^*\simeq\sqrt{2\alpha(t_0-t)}~.}
As $t\to t_0$, $\phi\to\phi^*$, and the time derivatives of $\phi$ diverge. The continuation of the trajectory past the singularity is not unique, since the solution has a branch cut starting at that point (but, see appendix D).

Some comments are useful at this point:
\item{(1)} The singularity of the trajectories we found occurs for any $E>R/l_s^2$. A natural interpretation of its appearance is that such trajectories do not give well behaved states in the quantum theory. This is in agreement with the expectations from the boundary theory described in the beginning of this section. We discuss the relation between the singularity of the trajectories and the upper bound on energies in appendix D.
\item{(2)} The location of the singularity, $\phi=\phi^*$ \phistar, depends on the energy of the probe string. As $E$ approaches $R/l_s^2$ (from above), $\phi^*\to\infty$. On the other hand, for large $E$, $\phi^*\to 0$, \ie\ the singularity is very close to the naked singularity at $r=r_1$ (just past it).
\item{(3)} The fact that the singularity of the trajectories occurs beyond the naive naked singularity in the geometry is in qualitative agreement with the expectation that we need to include the region beyond the singularity to describe string theory in $\MM_3^-$.
\item{(4)} The generalization of this section to string probe dynamics in the background of a BH in $\MM_3^\pm$ is presented in appendix E.
\item{(5)} The singularity of the trajectory of a probe string for $E>R/l_s^2$ has an interesting interpretation in terms of the induced metric on the string,
\eqn\indmet{ds^2={1\over f_1}\left[-dt^2(1-f_1\dot\phi^2)+dx^2\right]={\sqrt{1-f_1\dot\phi^2}\over f_1}\left[-dt^2\sqrt{1-f_1\dot\phi^2}+{dx^2\over\sqrt{1-f_1\dot\phi^2}} \right].}
Equation \energyyy\ implies that
\eqn\lglg{\sqrt{1-f_1\dot\phi^2}={R\over R+l_s^2 f_1E}.}
Plugging this into \indmet, and recalling that $f_1=-1+e^{-{2\phi\over\sqrt kl_s}}$ in $\MM_3^-$,
we see that the induced metric on the string looks near the singularity of the trajectories like the metric near the horizon of a BH (up to a conformal factor, that diverges there).

\newsec{Discussion}

In this note, we continued the study of string theory in the backgrounds $\MM_3^+$ \mthree, \fone, and $\MM_3^-$ \fonet, which correspond to the current-current deformation \deltal\ of string theory on $AdS_3$ with positive and negative coupling, respectively. We showed that the thermodynamics of black holes in $\MM_3^\pm$, \nzero, \mtmft, provides new insights into the dynamics of the theory.

The main focus of our discussion was on $\MM_3^-$ and its excitations. The background $\MM_3^-$ approaches $AdS_3$ in the infrared $(r\to 0)$. As we move out in the radial direction, the geometry changes, and at a finite value $r=r_1$, given in \fone, we encounter a singularity of the metric and dilaton. A study of excitations of the geometry reveals that this singularity does not constitute a hard wall in the theory.

In particular, when we consider (as we did in section 4) a black hole in this background, the location of the horizon of the BH, $r_0$, depends on its energy as in \ronem, which grows with the energy and in fact diverges when the energy approaches a critical value, \ec. Thus, the radial position of the horizon of the BH, $r_0$, can be much larger than $r_1$ \fone. However, we found that in the BH background, the position of the singularity is pushed to larger $r$ as well, $r=r_1(\EE)$ given in \ronem, and the singularity is always outside the horizon of the BH, $r_0(\EE)<r_1(\EE)$. In particular, the position of the singularity, $r_1(\EE)$, also diverges in the limit $\EE\to\EE_c$, and it is only in this limit that the horizon approaches the UV singularity.

We also studied (in section 5) the dynamics of a probe fundamental string in $\MM_3^\pm$. In $\MM_3^+$, we found that the trajectories are smooth, in agreement with the fact that the quantum theory contains a continuum of long string states starting from the supersymmetric vacuum (massless BTZ). In $\MM_3^-$, we found a richer picture: for energy below a critical value, $0<E<R/l_s^2$, we found smooth trajectories, that continue smoothly through the naked singularity, while for $E>R/l_s^2$, we found that the resulting trajectories are singular, with a singularity at a value of the radial coordinate that depends on the energy, but is always larger than $r_1$. We interpreted this as the classical analog of the fact that the quantum theory does not have states with $E>R/l_s^2$ (see appendix D).

One conclusion from these calculations is that the singularity at $r=r_1$ is soft -- when we add energy to the system, it occurs at a larger value of the radial coordinate. Another conclusion is that we cannot restrict attention to the region $r<r_1(0)$, since excitations like probe fundamental strings go smoothly through the singularity. Beyond the singularity, the signature of the time coordinate $t$ and spatial coordinate $x$ (the boundary coordinates of the original undeformed $AdS_3$) flips, but we argued that one should still think of $t$ as time and of $x$ as space.

The main motivation for this came from an analysis of the BH in section 4. We used standard techniques of black hole thermodynamics to determine the entropy, temperature and energy of this BH \entm, \hawt, \eaextm, and then asked whether we can reproduce these results by an ADM-type analysis on the boundary at infinity, beyond the singularity. We found that to do this we need to view the time coordinate $t$, which is spacelike near the boundary, as the time in the ADM construction, and discussed a potential interpretation of this fact (see appendix C).

The results of our analysis are in agreement with some known facts about string theory on $AdS_3$. As we mentioned in the introduction, the dynamics of long strings on $AdS_3$ is known to be holographically dual to a symmetric product of $p$ CFT's with central charge $c=6k$ \refs{\ArgurioTB,\GiveonMI}; see \ChakrabortyMDF\ for a recent discussion and further references. In this sector of the theory, the deformation \deltal\ can be shown to be a $T\bar T$ deformation of the block of the symmetric product.

A probe fundamental string can be thought of in the boundary theory as a state in a single block (symmetrized over all blocks). Thus, for negative coupling in \deltal, we expect to find a bound on the energies of such states, which is precisely the bound in $T\bar T$ deformed CFT with negative coupling. This explains the results of section 5 mentioned above.

A black hole corresponds in this language to a state in which all $p$ copies in the symmetric product are excited. Since the theory in each block of the symmetric product has a positive specific heat, in the large $p$ limit the energy is roughly equally divided among the $p$ factors. This explains the agreement of the BH thermodynamic analysis with the properties of a symmetric product of $T\bar T$ deformed CFT's.

It is interesting to compare our results to those of others that worked on related problems. In particular, the authors of \GuicaNZM\ discussed the holographic dual of $T\bar T$ deformation of a CFT with an $AdS_3$ dual in the gravity regime. They concluded that the picture of \McGoughLOL\ described earlier, with a UV wall at a fixed value of the radial coordinate, needs to be modified to include the whole region up to the boundary of $AdS_3$. This conclusion is in agreement with our results for the single-trace $T\bar T$ deformation -- as we discussed, the UV wall is soft, and excitations can smoothly pass through it. It would be interesting to pursue the analogy between the pictures of \GuicaNZM\ and this paper further.

We saw in section 4 that the theory with negative coupling can be thought of as describing $k$ NS5-branes and $p$ negative strings (see \plplpl). This resonates with the results of \CavagliaODA, who showed that the $T\bar T$ deformation of a free field theory of $D$ scalar fields gives for negative coupling a Nambu-Goto action for a string whose oscillator excitations contribute with a minus sign to the energy. A similar picture is obtained from comparing the exact spectrum of energies found in \refs{\SmirnovLQW,\CavagliaODA} to the spectrum of a (non-critical) free string (see \CallebautOMT\ for a recent discussion). It would be interesting to understand the relation between the two pictures.

Negative branes were also discussed in \DijkgraafLYM. These authors emphasized the fact that in string theory, negative branes are surrounded by a region with different signature metric than that obtained for large distance from the branes, and that the interface between the two regions can in some cases be regular, in the sense that probe branes can smoothly traverse it. Both of these elements are present in our example. Another point of contact between the two analyses is that the backgrounds with negative branes can be supersymmetric, and in fact preserve the same supersymmetry as the ones with positive branes. In our case, the backgrounds $\MM_3^\pm\times S^3\times T^4$ corresponding to $k$ NS5-branes and $p$ positive and negative strings, respectively, preserve the same $(4,4)$ supersymmetry.

It would be interesting to explore the connection between the work of \DijkgraafLYM\ and this note, and in particular see whether our results have implications for other systems, such as the ones studied there; see also \BlairOPS\ for a related idea. An example is the system of $N$ negative $D3$-branes, which is related to a
single-trace irrelevant deformation of $\NN=4$ SYM that preserves $\NN=4$ SUSY. Depending on the sign of the coupling, it is described in the bulk by a metric in which the harmonic function of $D3$-branes goes at large distances to $\pm 1$. In \ChakrabortyNME, we generalize our discussion of black holes to this case.

The region in $\MM_3^-$, which from our perspective is beyond the singularity, was discussed in a different context in~\JohnsonQT. In that paper, it was proposed that this ``repulson'' singularity is resolved by the ``enhancon'' mechanism. Many of the detailed features of the analysis are different in the two cases. For example, in \JohnsonQT, an important role was played by the fact that the fivebranes carried non-zero string charge. This is not the case in our system. Also, the physics studied in \JohnsonQT\ was associated with the region $r>r_1$, while in our case it is associated with the region $r<r_1$. As we discussed earlier, this is an important distinction. A related fact is that in~\JohnsonQT, the analog of our coordinate $x$ is non-compact. Nevertheless, since the two geometries look closely related, it is possible that one can learn from one about the other, and it would be interesting to do so.

There are many other open problems related to our work, some of which we list next.
\item{(1)} In \refs{\AharonyBAD,\AharonyICS}, it was shown that modular invariance of the target-space theory on a torus places very strong constraints on $T\bar T$ deformed CFT's and related theories. For $T\bar T$ deformed CFT with negative coupling this led to a clash between unitarity and modular invariance. The former favored keeping only states with real energies, while the latter did not allow such a truncation of the spectrum. The results of this note seem to suggest that a theory with the truncated spectrum may exist, and it would be interesting to understand what are its properties when placed on a torus.
\item{(2)}
The worldsheet sigma models on $\MM_3^\pm$ can be described by exact coset CFT's, obtained by gauging a null current on $SL(2,\IR)\times R^{1,1}$ \GiveonMYJ. Axial (vector) gauging gives $\MM_3^+$ $(\MM_3^-)$. The black hole in $\MM_3^+$ has a similar description, as an axially gauged ${SL(2,\IR)\times U(1)\over U(1)}$ \refs{\HorneGN,\GiveonMI}. The BH in $\MM_3^-$ can be obtained by a continuation $r_0^2\to -r_0^2$ of a vector gauging of ${SL(2,\IR)\times U(1)\over U(1)}$. CFT's in this class exhibit interesting non-perturbative effects in $l_s$ -- see \eg~\refs{\KutasovRR\GiveonDXE\GiveonGFK-\ItzhakiCGG} and references therein. It would be interesting to investigate their role in the present context. In particular, it would be interesting to understand if the $p$ negative strings that form the $\MM_3^-$ background are related to the order $1/g_s^2$ instant folded strings in the interior of the two-dimensional black hole discussed in~\refs{\ItzhakiGLF\AttaliGOQ-\GiveonXXH}.
\item{(3)}
The holographic duality between single-trace $T\bar T$ deformed $CFT_2$ and the worldsheet deformation \deltal\ is a special case of a more general duality between symmetric product CFT deformed by a general linear combination of single-trace $T\bar T$, $J\bar T$ and $T\bar J$, and perturbative string theory on backgrounds obtained by current-current deformations of $SL(2,\IR)\times U(1)$ \refs{\ChakrabortyVJA,\ChakrabortyMDF}. It was shown that in all these cases, the energies of states develop an imaginary part above certain critical values if and only if the dual geometries have pathologies such as singularities, closed timelike curves (CTC), and spacetime signature change beyond a certain value of the radial direction. For instance, the dual geometry to single-trace $\mu J\bar T$ deformed $CFT_2$ is the simplest version of a warped $AdS_3$ \ChakrabortyVJA, which has CTC's beyond a radial distance of order $1/|\mu|$ \ChakrabortyMDF, and the energy spectrum has a corresponding upper bound. It would be interesting to generalize our analysis of black holes and fundamental string probes to this background, and more generally to the whole family of backgrounds discussed in~\ChakrabortyMDF.
\item{(4)}
The properties of string theory on $\MM_3^-$, studied in this paper, as well as on warped $AdS_3$ and other examples mentioned in the previous item, may be of relevance to other backgrounds in quantum gravity with a finite number of states, such as de Sitter spacetime. It would be interesting to generalize our results to such spacetimes.

\bigskip\bigskip
\noindent{\bf Acknowledgements:}
We thank N. Itzhaki for collaboration on parts of this work, and O. Aharony, R. Argurio, S. Datta and A. Hashimoto for discussions. We also thank O. Aharony for comments on the manuscript. The work of SC is supported by the Infosys Endowment for the study of the Quantum Structure of Spacetime.
The work of AG and DK is supported in part by BSF grant number 2018068.
The work of AG is also supported in part by a center of excellence supported
by the Israel Science Foundation (grant number 2289/18).
The work of DK is also supported in part by DOE grant de-sc0009924.

\appendix{A}{BTZ black hole}

In section 3, we considered the near-extremal  geometry of $k$ NS5-branes wrapping $T^4\times S^1$, in the presence of $p$ fundamental strings wrapping the $S^1$ and carrying momentum $n$ along it. We recalled that in the near-horizon region of the fivebranes it can be thought of as a (rotating) charged black string in $\MM_3$ (times $S^3\times T^4$), with charge $p$ and angular momentum $n$. In this appendix, we show that in the near fundamental strings limit, one obtains a BTZ black hole.

As discussed in section 3, when $r_0\ll r_1$, the horizon of the BH is deep inside the $AdS_3$ region in $\MM_3^+$,
and taking the limit $r\ll r_1$, keeping $X$ in \X\ finite, amounts to dropping the $1$ from the harmonic function $f_1$
in \background, in addition to dropping the $1$ from $f_5$ (as already done in \ffive).
After a coordinate transformation, $\varphi=x/R$, $\tau=\sqrt{k}l_s t/R$,
and $\rho^2=\frac{R^2}{r_0^2\sinh^2\alpha_1}(r^2+r_0^2\sinh^2\alpha_n)$,
the metric takes the form
\eqn\btz{ds^2=-N^2d\tau^2+\frac{d\rho^2}{N^2}+\rho^2(d\varphi+N_\varphi d\tau)^2,}
with
\eqn\nfunc{\eqalign{N^2=\frac{(\rho^2-\rho_+^2)(\rho^2-\rho_-^2)}{l_s^2k\rho^2}
=\frac{\rho^2}{l_s^2k}-{8G_3M}+\frac{(4G_3n)^2}{\rho^2},\qquad
N_{\varphi}=-\frac{4G_3n}{\rho^2},}}
where the three dimensional Newton constant, $G_3$, the location of the outer and inner horizons, $\rho_\pm$,
mass $M$ and angular momentum $n$,
are given in terms of the parameters in section 2 by
\eqn\gnthree{G_3={e^{2\Phi(r\to 0)}l_s\over 4vk^{3/2}}=\frac{l_s^3}{4v\sqrt{k}X\sinh^2\alpha_1},}
\eqn\horizon{\rho_+^2=\frac{\cosh^2\alpha_n}{\sinh^2\alpha_1}R^2~,\qquad \rho_-^2=\frac{\sinh^2\alpha_n}{\cosh^2\alpha_1}R^2~,}
\eqn\mj{{\sqrt k l_s}M=\frac{R^2vX}{2l_s^4}\cosh 2\alpha_n=h+\bar{h}-\frac{c}{12}~,\qquad n=\frac{R^2vX}{2l_s^4}\sinh 2\alpha_n=h-\bar{h}~.}
The geometry thus obtained is that of a BTZ black hole,\foot{For a review, see \eg\ \CarlipZN\ and references therein.}
and in eq. \mj\ we indicated the value of its mass and angular momentum
within the standard $AdS_3/CFT_2$ dictionary, where $h$ and $\bar{h}$ are left and right handed scaling dimensions
in a $CFT_2$ with central charge $c=6kp$.

\appendix{B}{General $n$}
In this appendix, we present the result for the entropy and spectrum of black holes in $\MM_3^+$, for generic momentum on the $S^1$, $n/R$. From eq. \sthree, in the limit \ffive, one finds the familiar result \GiveonMI, that the entropy of a black string which consists of $p$ fundamental strings wrapping the $S^1$ with momentum $n$, in the near-horizon region of $k$ NS5-branes wrapping $T^4\times S^1$, is
\eqn\entgn{S=\frac{\pi R r_0^2\cosh\alpha_1\cosh\alpha_n}{2G_{N}^{(3)}kl_s^2}=
\pi \sqrt{k}l_s\left(\sqrt{m^2-q_L^2}+\sqrt{m^2-q_R^2}\right),}
where the mass $m$ and charges $q_{L,R}$ are given in terms of the winding and momentum numbers, $p,n$,
and the dimensionless energy above extremality, ${\cal E}$, by
\eqn\mqlqr{m=\frac{pR}{l_s^2}+{{\cal E}\over R}~, \ \ \ \ q_{L,R}=\frac{n}{R}\pm \frac{pR}{l_s^2}~.}
As in section 3, we regard $\lambda\equiv{l_s^2/R^2}$ as a parameter that we change while keeping $S$ fixed.
In particular, in the $\lambda\to 0$ limit, we find that $S$ in \entgn\ is indeed the entropy of a BTZ black hole,
\btz\ -- \mj, with mass and angular momentum $M=\EE(0)/\sqrt k l_s$ and $n$, respectively,
\eqn\entzero{S=2\pi \sqrt{kp\over 2}\left(\sqrt{{\cal E}(0)-n}+\sqrt{{\cal E}(0)+n}\right)
=2\pi\sqrt{c\over 6}\left(\sqrt{h-{c\over 24}}+\sqrt{\bar h-{c\over 24}}\right)~,}
as it should.
And, finally, from \entgn\ = \entzero, one finds that the energies deform as
\eqn\geneng{\frac{1}{p}{\cal E}(\lambda)=\frac{1}{\lambda}\left(-1+\sqrt{1+2\lambda\frac{{\cal E}(0)}{p}+\lambda^2\left(\frac{n}{p}\right)^2}\right),}
which is precisely the spectrum found in \refs{\SmirnovLQW,\CavagliaODA},
written in terms of the energy and momentum per block, $\EE/p$ and $n/p$,
in a symmetric product of $p$ $T\bar T$ deformed CFT's.

\appendix{C}{ADM mass}

In this appendix, we will present a brief derivation of the ADM mass of the the black holes discussed at various stages in the paper.
We start with the background \background. The metric in the Einstein frame is given by
\eqn\eframe{ds^2_E= e^{-(\Phi-\Phi_0)/2}ds^2.}
Dimensionally reducing the metric on $T^4\times S^1$ to five dimensions gives rise to a $4+1$ dimensional black hole
\eqn\fdbh{ds_E^2= -(f_1f_5f_n)^{-2/3}f dt^2+(f_1f_5f_n)^{1/3}\left(\frac{dr^2}{f}+r^2d\Omega^2_3\right), }
where $f$ and $f_{1,5,n}$  are given by \harfun, as before.
According to the ADM prescription (see \eg\ \MyersYC\ for a review), the mass of the BH can be read off from the asymptotic expansion of the time component of the metric
\eqn\gtt{g_{tt}=-\left(1-\frac{4 G_{N}^{(10)}}{3 \pi^2 R V_{T^4}}\frac{M_{\rm ADM}}{r^2}+O(1/r^4)\right).}
Comparing \gtt\ to the asymptotic expansion of the time component of the metric in \fdbh\ gives rise to \mADM.

Next, we consider the background \background\ with $\alpha_n=0$,  $f_1=-1+r_0^2\cosh^2\alpha_1/r^2$ and $f_5$ given by \harfun. The metric in the Einstein frame is again given by \eframe. Since $f_1$ \fonet\ is negative for $r>r_1$, the dilaton in \background\ appears to be complex. We take the attitude\foot{This can be justified by studying the corresponding worldsheet sigma model.} that in the expression for the dilaton we need to replace $f_1$ by $|f_1|$. Upon dimensional reduction to $4+1$ dimensions, one obtains
\eqn\fdbhb{ds_E^2= -\frac{|f_1|^{1/3}}{f_1}f_5^{-2/3}f dt^2+(|f_1|f_5)^{1/3}\left(\frac{dr^2}{f}+r^2d\Omega^2_3\right),}
which describes a black hole with a horizon at $r=r_0$ and a naked singularity at $r=r_1$.

As discussed in section 4, $t$ is spacelike in the asymptotically flat region $r\to\infty$, whereas $x$ is timelike there. Thus, to compute the ADM mass of the resulting spacetime, one might think that $x$ should be treated as time. However, it turns out that to match to the thermodynamic discussion of section 4 one has to still treat $t$ as time, and to define the ADM mass
by the following asymptotic expansion of the metric component $g_{tt}$:
\eqn\gttb{g_{tt}=\left(1-\frac{2 G_{N}^{(10)}}{3 \pi^2 R V_{T^4}}\frac{M_{\rm ADM}}{r^2}+O(1/r^4)\right).}
Indeed, comparing \gttb\ to the asymptotic expansion of the corresponding component of the metric in \fdbhb, one obtains
\eqn\badm{M_{\rm ADM}= \frac{R vX}{2l_s^4}(-\cosh2\alpha_1+\cosh2\alpha_5+1).}
In the limit $r_0\to 0$, \badm\ takes the form
\eqn\bext{E_{\rm ext}=-p{R\over l_s^2}+k{\frac{R v}{g^2l_s^2}}=\frac{R vX}{2l_s^4}(-\sinh2\alpha_1+\sinh2\alpha_5),
}
which looks like the energy of $p$ negative strings and $k$ fivebranes. In the decoupling limit of section 3, the energy $E=M_{\rm ADM}-E_{\rm ext}$ takes the form \eaextm.

At first sight the above prescription seems strange, but we want to argue that it is analogous to what we do in many other settings in QFT. For example, to study quantum electrodynamics in an external electric field in Lorentzian signature, we turn on a non-zero expectation value of $F_{01}$ (say), the electric field in the $x$ direction. Now suppose we want to study the theory in Euclidean signature (\eg\ in order to analyze it at finite temperature). Since under Wick rotation $t\to i\tau$ the field strength $F_{01}\to iF_{01}$, the resulting Euclidean theory has an imaginary condensate of the field strength. This looks odd from the point of view of the Euclidean theory, but is OK since we are using the Euclidean theory to learn about the Lorentzian one.

In our case, the situation is similar, except the flip of signature happens as a function of the radial coordinate, rather than being done by hand. Since we are interested in the dynamics of the region $r<r_1$, where the time is $t$, it is natural to perform the ADM procedure with respect to that time, even though it is no longer timelike near the boundary at infinity, where the ADM analysis is performed.

\appendix{D}{Semiclassical analysis}

In section 5 we saw that a probe string with energy $E>R/l_s^2$ in $\MM_3^-$ encounters a singularity along its trajectory when it reaches the point $\phi^*$ \phistar.  We also argued (in comment (1) after eq. \solphi) that this is related to the fact that states with these energies develop an imaginary part in their energy. The purpose of this appendix is to explore this relation.

The analysis of section 5 involved the Nambu-Goto action for the string (with a coupling of the string to the external B-field, given by the second term in \laggg). To study the trajectory of the string we fixed the static gauge and solved for the trajectory of the string in the radial direction, $\phi=\phi(t)$, \phidottt.

As is well known, for quantization it is more convenient to go from the Nambu-Goto action to the Polyakov one. The Nambu-Goto action is obtained by solving the equations of motion for the worldsheet metric $h_{ab}$, which set it equal (up to an undetermined Weyl factor) to the induced metric on the string. That metric is gven by \indmet, and was discussed in section 5.

For quantization, it is more convenient to fix the conformal gauge $h_{ab}=\eta_{ab}$ (again, up to a conformal factor). In this appendix we will use the resulting action to do two things: (1) review the (semiclassical) derivation of the energy formula and in particular the resulting imaginary part above the critical energy; (2) describe the classical trajectories of section 5, whose semiclassical quantization gives those energies. Doing both in conformal gauge makes the relation between them more or less manifest.

The worldsheet action of a fundamental string in $\MM_3^\pm$ in conformal gauge is given by
\eqn\slag{S=\frac{1}{2\pi l_s^2}\int d^2z \left(\partial\phi\bar{\partial}\phi+{1\over f_1}\partial\bar{\gamma}\bar{\partial}\gamma\right),}
where the worldsheet is a cylinder parametrized by $\tau$ and $\sigma\simeq \sigma+2\pi$, with
\eqn\zzb{z=\frac{1}{\sqrt{2}}(\tau+\sigma), \ \ \ \ \bar{z}=\frac{1}{\sqrt{2}}(\tau-\sigma)}
and
\eqn\der{\partial=\frac{1}{\sqrt{2}}(\partial_\tau+\partial_\sigma),\ \ \ \ \bar{\partial}=\frac{1}{\sqrt{2}}(\partial_\tau-\partial_\sigma).}
The coordinates $(\gamma,\bar{\gamma})$ in \slag\ are related to $(t,x)$ in section 5 by
\eqn\ggbp{\gamma=x+t, \ \ \ \bar{\gamma}=x-t.}

\bigskip

\noindent {D.1. {\it Semiclassical quantization }}

The conjugate momentum densities for the fields $(\phi,\gamma,\bar{\gamma})$ are given by
 \eqn\momcon{\eqalign{
&\Pi_{\phi}=T\dot{\phi},\cr
&\Pi_{\gamma}=\frac{T}{\sqrt{2}}f_1^{-1}\partial\bar{\gamma}=\frac{T}{2}f_1^{-1}(\dot{\bar{\gamma}}+\bar{\gamma}'),\cr
&\Pi_{\bar{\gamma}}=\frac{T}{\sqrt{2}}f_1^{-1}\bar{\partial}\gamma =\frac{T}{2}f_1^{-1}(\dot{\gamma}-\gamma'),}}
where $T$ is the string tension given by $T=1/2\pi l_s^2$, and dot and prime denote derivatives with respect to $\tau$ and $\sigma$ respectively.

The worldsheet Hamiltonian is given by
\eqn\wsham{H=\frac{1}{2\pi}\int_0^{2\pi}d\sigma \cal{H},}
where the Hamiltonian density is
 \eqn\hamd{\eqalign{\cal{H}&=\Pi_{\phi}\dot{\phi}+\Pi_{\gamma}\dot{\gamma}+\Pi_{\bar{\gamma}}\dot{\bar{\gamma}}-\cal{L}\cr
 &= \Pi_{\gamma}\gamma'-\Pi_{\bar{\gamma}}\bar{\gamma}'+\frac{2\Pi_{\gamma}\Pi_{\bar{\gamma}}f_1}{T}+\frac{\Pi_{\phi}^2}{2T}+\frac{T\phi'^2}{2}.}}
The worldsheet momentum is given by
\eqn\wsmom{p=\frac{1}{2\pi}\int_0^{2\pi}d\sigma \cal{P},}
where the worldsheet momentum density $\cal{P}$ is
\eqn\wsmomd{{\cal{P}} =\Pi_{\phi}\phi'+\Pi_{\gamma}\gamma'+\Pi_{\bar{\gamma}}\bar{\gamma}'.}
Let $E$ be the energy of the string conjugate to $t$ and $P$ be the momentum conjugate to $x$. Since $x$ is compact, $P$ is quantized
\eqn\momp{P=\frac{n}{R},}
where $n$ is an integer.
 The spacetime quantum numbers are given by
\eqn\stqn{\eqalign{
&\int_0^{2\pi}d\sigma\Pi_{\gamma}=-\frac{E_L}{R},\cr
&\int_0^{2\pi}d\sigma\Pi_{\bar{\gamma}}=\frac{E_R}{R},\cr
&\int_0^{2\pi}d\sigma\Pi_{\phi}=P_{\phi},}}
where
\eqn\enqn{\eqalign{
&E_L=\frac{R}{2}(E+P),\cr
&E_R=\frac{R}{2}(E-P).}}
We also have
\eqn\stqnn{\eqalign{
&\int_0^{2\pi}d\sigma \gamma'=\int_0^{2\pi}d\sigma\bar{\gamma}'=2\pi w R,\cr
&\int_0^{2\pi}d\sigma\phi'=0,}}
where $w$ is the winding of the string around the $x$ circle. We will mainly be interested in $w=1$, since this is the sector which can be most directly compared to $T\bar T$ deformed CFT.

The contribution of the zero modes of the string to the worldsheet Hamiltonian $H$, denoted by $\Delta+\bar{\Delta}$, is given by
\eqn\del{\Delta+\bar{\Delta}=-w(E_L+E_R)-\frac{l_s^2f_1}{2}(E^2-P^2)+\frac{l_s^2}{2}P_{\phi}^2~.}
The contribution of the zero modes of the string to the worldsheet momentum $p$, is given by
\eqn\dell{\bar{\Delta}-\Delta=nw.}
For a string with zero momentum in the $x$ direction, $P=n=0$, winding $w=1$, and no excitation of the transverse directions, the vanishing of the total stress tensors, $T=\bar T=0$ (the Virasoro constraints), implies that $\Delta=\bar\Delta=0$ in \del. The resulting equation depends on $\phi$. This is due to the fact that the worldsheet theory \slag\ is interacting. As usual in theories of this sort (\eg\ Liouville theory, CFT on the cigar, etc), it is convenient to send $\phi\to\infty$, where $f_1$ approaches a constant, the worldsheet theory becomes free, and $P_\phi$ becomes a good quantum number.

The solution of the mass-shell condition is
\eqn\semimass{E={R\over l_s^2}-\sqrt{\left(R\over l_s^2\right)^2-P_\phi^2}}
We see that when the momentum $P_\phi$ exceeds the critical value $P_\phi=R/l_s^2$, the energy develops an imaginary part. In terms of $T\bar T$ deformed CFT, one can think of $P_\phi$ as labeling the undeformed energy of the state corresponding to the long string, $E_0$. The precise relation is $E_0 R=\half\alpha' P_\phi^2$.

\bigskip

\noindent {D.2. {\it Classical trajectories}}

The purpose of this subsection is to find the classical trajectories that give rise upon semiclassical quantization to the states described by \semimass. We will find that they are precisely the trajectories discussed in section 5, viewed in conformal gauge. As in \semimass, we will restrict to states with $P=0$, for simplicity.

The equations of motion obtained by varying the action \slag\ are
\eqn\seom{\eqalign{
&2\partial\bar{\partial}\phi=-\frac{f_1'}{f_1^2}\partial\bar{\gamma}\bar{\partial}\gamma,\cr
& f_1^{-1}\bar{\partial}\gamma=c,\cr
&f_1^{-1}\partial\bar{\gamma}=-c,}}
where $f_1'$ denotes derivative of $f_1$ with respect to $\phi$ and
\eqn\const{c=\frac{E}{2\sqrt{2}\pi T},}
derived using \stqn.
To solve the equations of motion \seom, we choose the ansatz
\eqn\ansatz{
\phi=\phi(\tau), \ \ \ \ x=R \sigma, \ \ \ \ t =t(\tau).}
Then \seom\ takes the form
\eqn\aeom{\eqalign{
&\ddot{\phi}=\frac{f_1'}{2f_1^2}(\dot{t}-R)^2,\cr
&\dot{t}=R+l_s^2f_1E.}}
One can compute the stress tensor and imposing the Virasoro constraints get
\eqn\virc{\dot{\phi}^2-f_1^{-1}(\dot{t}^2-R^2)=0.}
Using \virc\ and the second equation in \aeom, one gets
\eqn\speed{\left(\frac{d\phi}{dt}\right)^2=\frac{l_s^2E(2R+l_s^2f_1E)}{(R+l_s^2f_1E)^2},}
which is precisely the result obtained in \phidottt.

As discussed in section 5, in $\MM_3^-$ \speed\ leads to singular trajectories for $E>R/l_s^2$. In conformal gauge, we can see this singularity by combining equations \aeom\ and \virc. For energy $E<R/l_s^2$, we can study trajectories for which $t(\tau),\phi(\tau)\to\infty$ at large worldsheet time $\tau$. For such trajectories, at large $\tau$ $f_1(\phi)\to -1$, and \aeom\ gives $t(\tau)\simeq (R-l_s^2 E)\tau$.

Substituting into \virc, and using the fact that $\dot\phi=l_s^2P_\phi$ (see \momcon, \stqn), we find that $\phi(\tau)\simeq l_s^2P_\phi\tau$, where $P_\phi$ is related to the energy as in \semimass. As $P_\phi$ increases, the energy $E$ \semimass\ approaches the critical value $R/l_s^2$. Increasing $P_\phi$ further does not increase the energy beyond the critical value, but rather introduces an imaginary part, which presumably means that the state is unstable.

This seems to be compatible with the fact that the classical trajectory \aeom, \virc, has the property that for $E>R/l_s^2$, $\dot t$ changes sign at a finite value of $\phi$. Thus, the trajectory $(t(\tau),\phi(\tau))$ cannot approach the boundary at infinity at large $\tau$. At the same time, substituting \aeom\ into \virc, one finds that $\phi$ goes to infinity at large $\tau$. Hence, this solution corresponds to a string that does not exist beyond some finite time.

\appendix{E}{String probe dynamics in $\MM_3$ with a black hole}

In this appendix, we generalize the discussion in section 5 to the case of a BH in $\MM_3^{\pm}$ (\ie\ to $r_0>0$).
The dynamics of a probe string in $\MM_3^{\pm}$ with $r_0>0$ is governed by the Lagrangian
\eqn\laggg{{\cal L}
=-\frac{1}{2\pi l_s^2f_1}\left(\sqrt{\left(f-\frac{f_1}{f}\dot{\phi}^2\right)}-f_1B_{tx}\right),}
where, as there, the first term in the brackets is the Nambu-Goto Lagrangian, while the second is the coupling of a fundamental string to the $NS$ B-field. The B-field in $\MM_3^{\pm}$ with $r_0>0$ is given by
\eqn\bfield{B_{tx}=\pm\left(-\frac{r_0^2\sinh2\alpha_1}{2r_1^2f_1e^{\frac{2\phi}{\sqrt{k}l_s}}}+1\right),}
and $f_1=\pm1+r_1^2/r^2$, with $r_1^2$ given in eqs. \harfun\ and \rrcc\ for the $+$ and $-$ cases, respectively,
and $\phi=\sqrt{k}l_s\ln(r/r_1)$.

Unlike $\MM_3^{\pm}$, when $r_0>0$, a probe string at rest feels a non-zero force. The corresponding potential is
\eqn\pote{V(\phi)=-\int_0^{2\pi R}dx \ {\cal L}(\dot{\phi}=0)= \frac{R(\sqrt{f}-f_1B_{tx})}{l_s^2 f_1}~.}
In both cases, the potential energy outside the horizon, $r>r_0$,
is a negative, monotonically increasing smooth function of $\phi$,
which approaches $V\to 0^-$ as $\phi\to\infty$.
In particular, $V$ is smooth at $r=r_1$,
even though the classical geometry has a curvature singularity there.

The canonical momentum of $\phi$ is given by
\eqn\pimom{\Pi=\frac{\delta {\cal L}}{\delta \dot{\phi}}=\frac{1}{2\pi l_s^2 f} \frac{k\dot{\phi}}{\sqrt{f-\frac{f_1}{f}\dot{\phi}^2}}~,}
and the energy of the string is
\eqn\enrgy{E=\int_0^{2\pi R} dx( \Pi \dot{\phi}-{\cal L})=\frac{R}{l_s^2f_1}\left(\frac{f}{\sqrt{f-\frac{f_1}{f}\dot{\phi}^2}}-f_1B_{tx}\right)~.}
Note that for a stationary string in $\MM_3^{\pm}$ with $r_0>0$, the energy no longer vanishes. This is due to the fact that the black hole breaks supersymmetry, so the string is no longer a BPS state.

One can solve \enrgy\ for the velocity of the string:
\eqn\vel{\dot{\phi}^2=\left(\frac{f^2}{f_1^3}\right) \frac{f_1^2(RB_{tx}+l_s^2E)^2-fR^2}{(RB_{tx}+l_s^2E)^2}.}
The dynamics of the probe string in the background of a black hole in $\MM_3^+$ is quite different from that in $\MM_3^-$.
Thus, we will treat the two cases separately.

\bigskip

\noindent {E.1. {\it Probe string in } $\MM_3^+$ {\it with } $r_0>0$}

As an example,
let us consider a probe string in the background $\MM_3^+$ with a BH whose horizon $r=r_0$ is located  deep in the $AdS_3$ regime.
Let us also consider the case that the string starts at $t=0$ and $\phi(t=0)=\phi_0$, where $\phi_0$ is very close to the horizon, but is outside the BH, and is moving in the radial outward direction with energy $E\ll R/l_s^2$. As it moves towards larger $\phi$, it accelerates, and for large positive $\phi$, in the region where $f_1(\phi)\simeq 1$, its velocity approaches a constant value, $\dot\phi\simeq l_s\sqrt{2E/R}$. It takes the string an infinite amount of time to reach the boundary, which is located at $\phi\to\infty$.
This behavior is quite similar to the case $r_0=0$.

Similar to pure $\MM_3^+$, the classical string described above gives rise in the quantum theory to a continuum of long string states labeled by the energy $E$. This energy takes values in $\IR_+$, and in particular there is no upper bound on it.
Things are very different for long strings in the background of a black hole in $\MM_3^-$, as we shall discuss next.

\bigskip

\noindent {E.2. {\it Probe string in } $\MM_3^-$ {\it with } $r_0>0$}

As in the case of pure $\MM_3^-$, $f_1$ here is positive for large negative $\phi$, but it goes to zero at the singularity, and then becomes negative, approaching $-1$ at large positive $\phi$. This leads to a difference in the resulting trajectories.
Inspecting the relevant equations above, one sees that the probe string is moving smoothly
in the regime between the horizon and singularity, $r_0<r<r_1$,
in particular, at the singularity.
In fact, the velocity $\dot\phi$ and all higher time derivatives of $\phi$ are finite at that point.
However,
continuing the trajectory of the string past the singularity, we see that there are two distinct regimes of energy. For $E<R/l_s^2$, the solution remains smooth for all $t$. At large $t$, $\phi\to\infty$, so $f_1\to -1$, and $\dot\phi$ in \vel\ approaches a constant value given by \phiass.
For $E>R/l_s^2$, the e.o.m. \vel\ has a singularity at a finite value of $\phi$,
\eqn\rstar{\phi^\ast=\frac{1}{2}\sqrt{k}l_s\log\left(1+\frac{R\sqrt{1-r_0^2/r_1^2}}{E l_s^2-R}\right)=\frac{1}{2}\sqrt{k}l_s \log\left(1+\frac{Rr_0^2\sinh2\alpha_1}{2r_1^2(E l_s^2-R)}\right).}
Near the singularity, it takes the form
\eqn\nrsig{\dot\phi\simeq-{\alpha\over\phi-\phi^*}~,}
where
\eqn\not{\alpha=\frac{l_s\sqrt{kR}r_0^2\sinh2\alpha_1
\left[2Rr_1^2+(El_s^2-R)r_0^2\sinh2\alpha_1\right]^{3/2}}{4\sqrt{2}r_1^3(El_s^2-R)\left[2r_1^2(El_s^2-R)+Rr_0^2\sinh2\alpha_1\right]}~,}
with the solution
\eqn\sphi{\phi-\phi^*\simeq\sqrt{2\alpha(t_0-t)}~.}
As $t\to t_0$, $\phi\to\phi^*$, and the derivatives of $\phi$ diverge.

As in the case of $\MM_3^-$ with $r_0=0$, the continuation of the trajectory past the singularity is not unique, since the solution has a branch cut starting at that point.  The properties of the probe string trajectory in $\MM_3^-$ with $r_0=0$, stated in comments (1), (2), (3) at the end of subsection 5.2, also hold here.

\listrefs
\end